\documentclass[conference]{IEEEtran}           % onecolumn (standard format)
\usepackage{graphics,subfigure}
\usepackage{epsfig}
\usepackage{array}
\usepackage[cmex10]{amsmath}
\usepackage{graphicx}
\usepackage{amssymb}
\usepackage{setspace,multirow}
\usepackage{multicol}
\usepackage{color, colortbl}
\usepackage{booktabs}
\usepackage{bigstrut}
\usepackage{floatrow}
\usepackage{rotating}
\usepackage{dblfloatfix} %To enable figures at the bottom of page
\usepackage[misc]{ifsym} %To insert symbol like email
\usepackage[leftcaption]{sidecap}
\graphicspath{{images/}}
\def\BibTeX{{\rm B\kern-.05em{\sc i\kern-.025em b}\kern-.08em
    T\kern-.1667em\lower.7ex\hbox{E}\kern-.125emX}}
\begin{document}

\title{A Residual Encoder Decoder Network for Segmentation of Retinal Image Based Exudates in Diabetic Retinopathy Screening\\
%{\footnotesize \textsuperscript{*}}
%\thanks{}
}
\author{Malik A. Manan$^1$,
Tariq M. Khan$^2$
Ahsan Saadat$^3$,
 Muhammad Arsalan$^1$, and
 Syed S. Naqvi$^1$\\
$^1$ Dept. of Electrical and Computer Eng., COMSATS University Islamabad,
Islamabad, Pakistan\\
$^2$ School of IT, Deakin University, Waurn Ponds, VIC 3216, Australia\\
$^3$ Victoria University, Sydney, Australia}

\maketitle

\begin{abstract}
Diabetic retinopathy refers to the pathology of the retina induced by diabetes and is one of the leading causes of preventable blindness in the world. Early detection of diabetic retinopathy is critical to avoid vision problem through continuous screening and treatment. In traditional clinical practice, the involved lesions are manually detected using photographs of the fundus. However, this task is cumbersome and time-consuming and requires intense effort due to the small size of lesion and low contrast of the images. Thus, computer-assisted diagnosis of diabetic retinopathy based on the detection of red lesions is actively being explored recently. In this paper, we present a convolutional neural network with residual skip connection for the segmentation of exudates in retinal images.
To improve the performance of network architecture, a suitable image augmentation technique is used. The proposed network can robustly segment exudates with high accuracy, which makes it suitable for diabetic retinopathy screening. Comparative performance analysis of three benchmark databases: HEI-MED, E-ophtha, and DiaretDB1 is presented. It is shown that the proposed method achieves accuracy (0.98, 0.99, 0.98) and sensitivity (0.97, 0.92, and 0.95) on E-ophtha, HEI-MED, and DiaReTDB1, respectively.
\end{abstract}
%\begin{keyword}
%Diabetic retinopathy\sep Semantic segmentation \sep Exudates \sep Machine learning \sep Deep neural network
%\end{keyword}

\section{\bf \textsc{Introduction}}

Recent studies have shown an increased trend in adults getting diagnosed with diabetes. Diabetes can cause many diseases in human body, such as kidney failure, heart issues, stroke and loss of vision. Diabetic retinopathy (DR) is a diabetic  complication that affects eyes and is the main cause of blindness in diabetes triggered diseases \cite{soomro2017contrast,soomro2017computerised,khan2019boosting,soomro2018impact}. Diabetic retinopathy occurs due to long-term cumulative damage of small retinal vessels in the retina. Early detection and treatment of diabetic retinopathy is necessary to prevent vision loss \cite{khan2019generalized,khawaja2019improved,khawaja2019multi}. Unfortunately, due to an increase in the number of people with diabetes and lack of trained professionals, especially in the under developed regions, diabetic retinopathy treatment is a public health problem worldwide \cite{khan2021residual}.

\begin{figure}[!t]
\centering
\subfigure[] {\label{cropping_a}\includegraphics[width=1.3in]{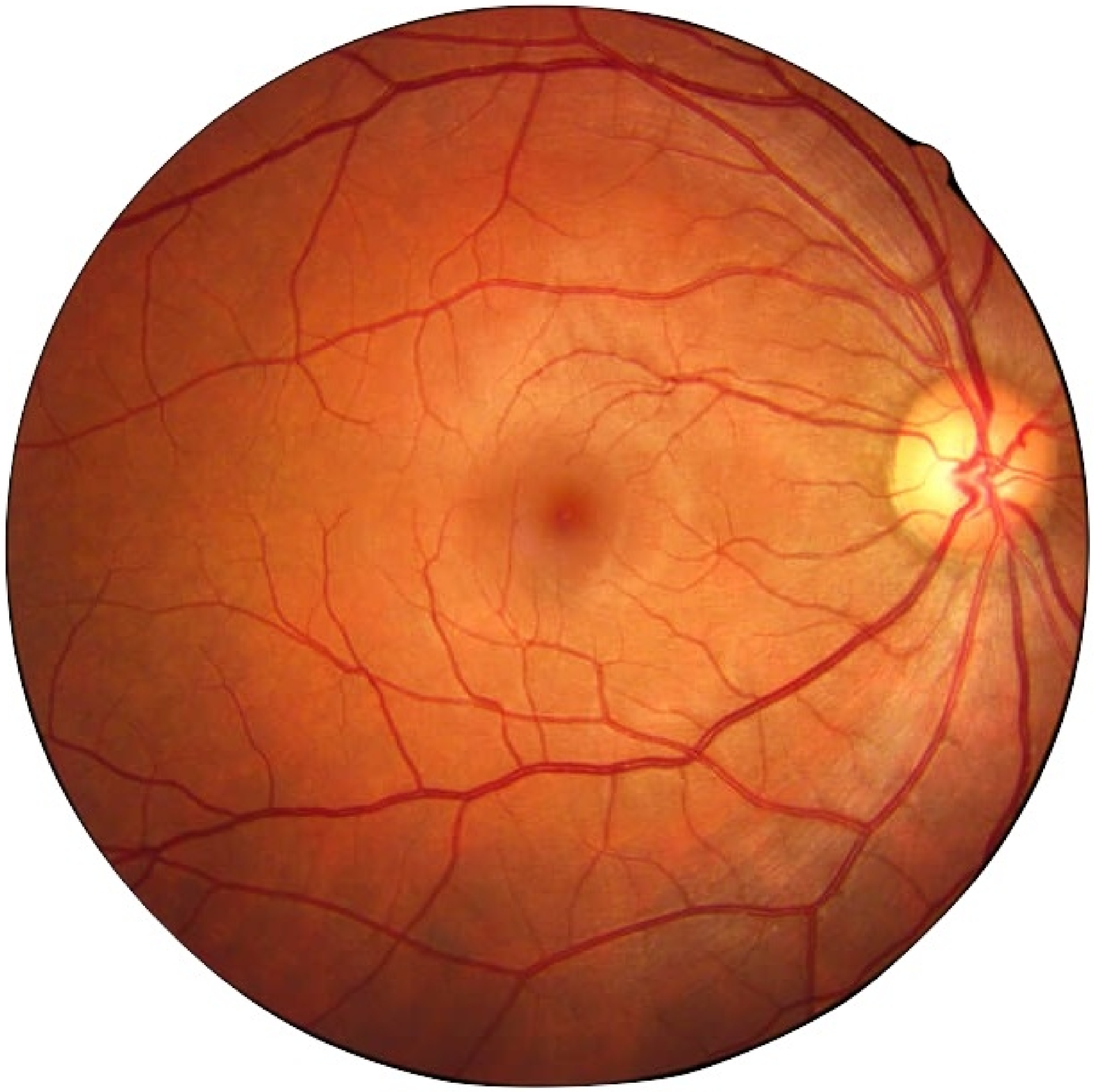}}
\subfigure[] {\label{cropping_b}\includegraphics[width=1.3in]{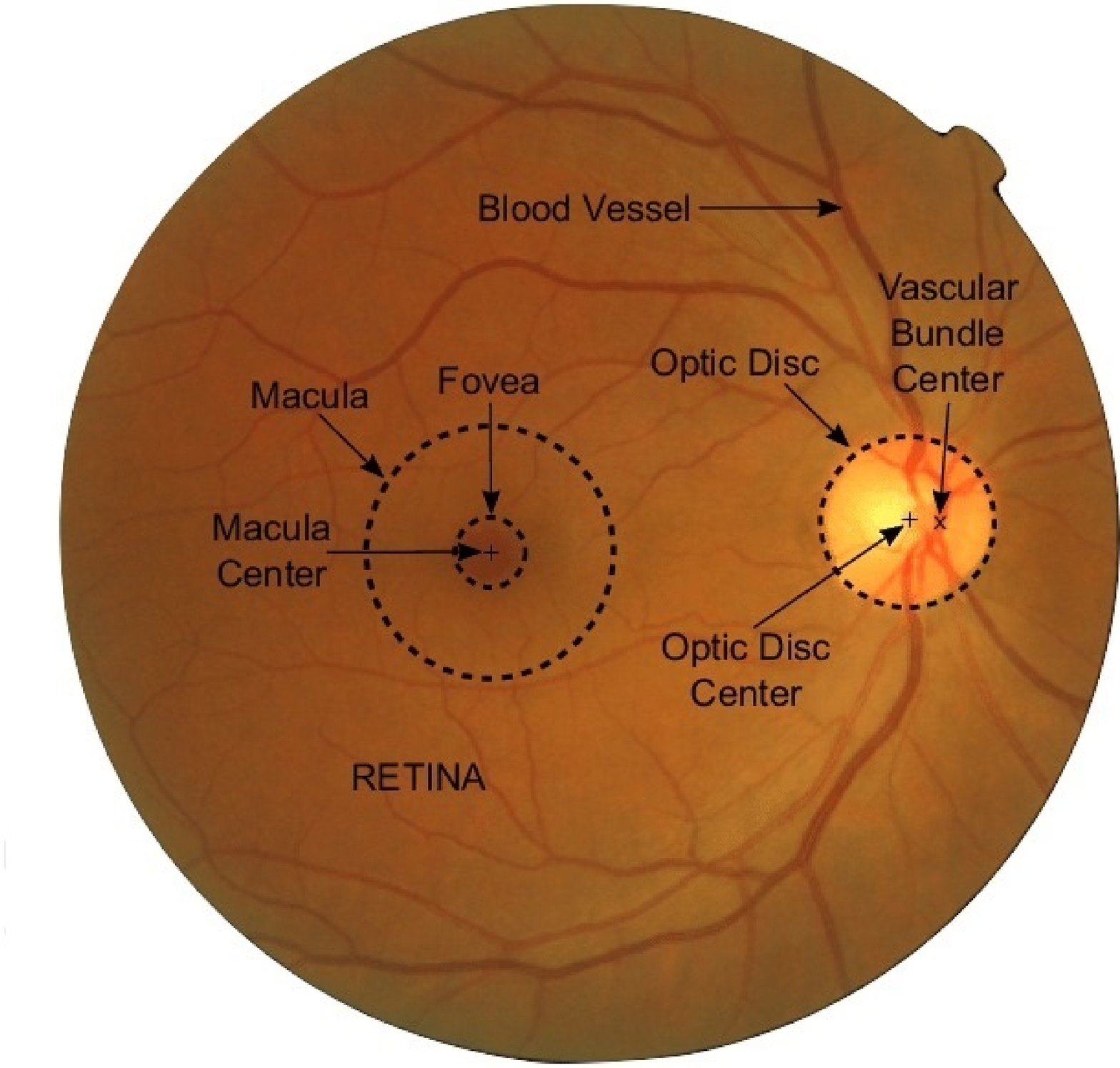}}
\subfigure[] {\label{cropping_a}\includegraphics[width=1.3in]{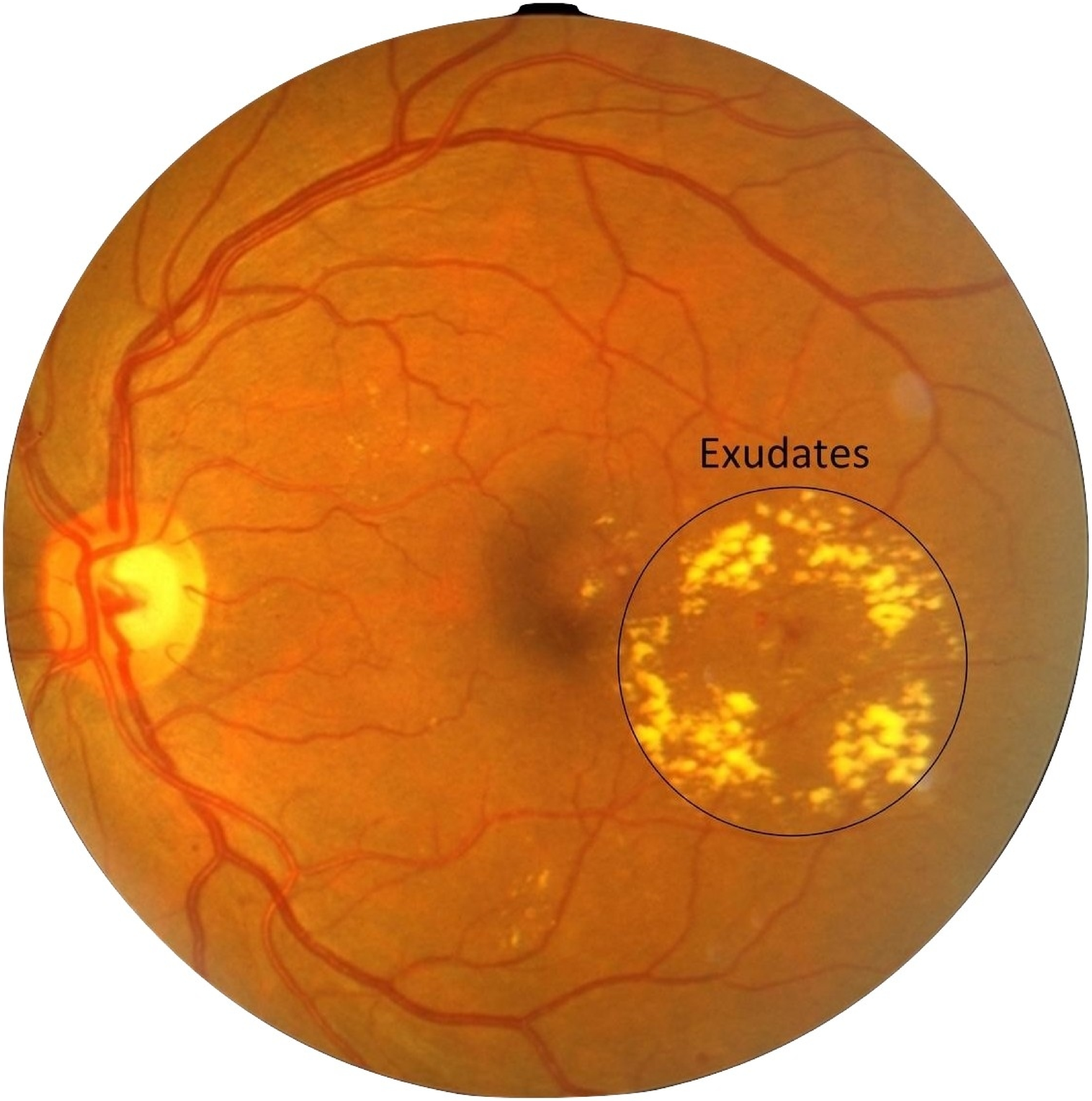}}
\subfigure[] {\label{cropping_a}\includegraphics[width=1.3in]{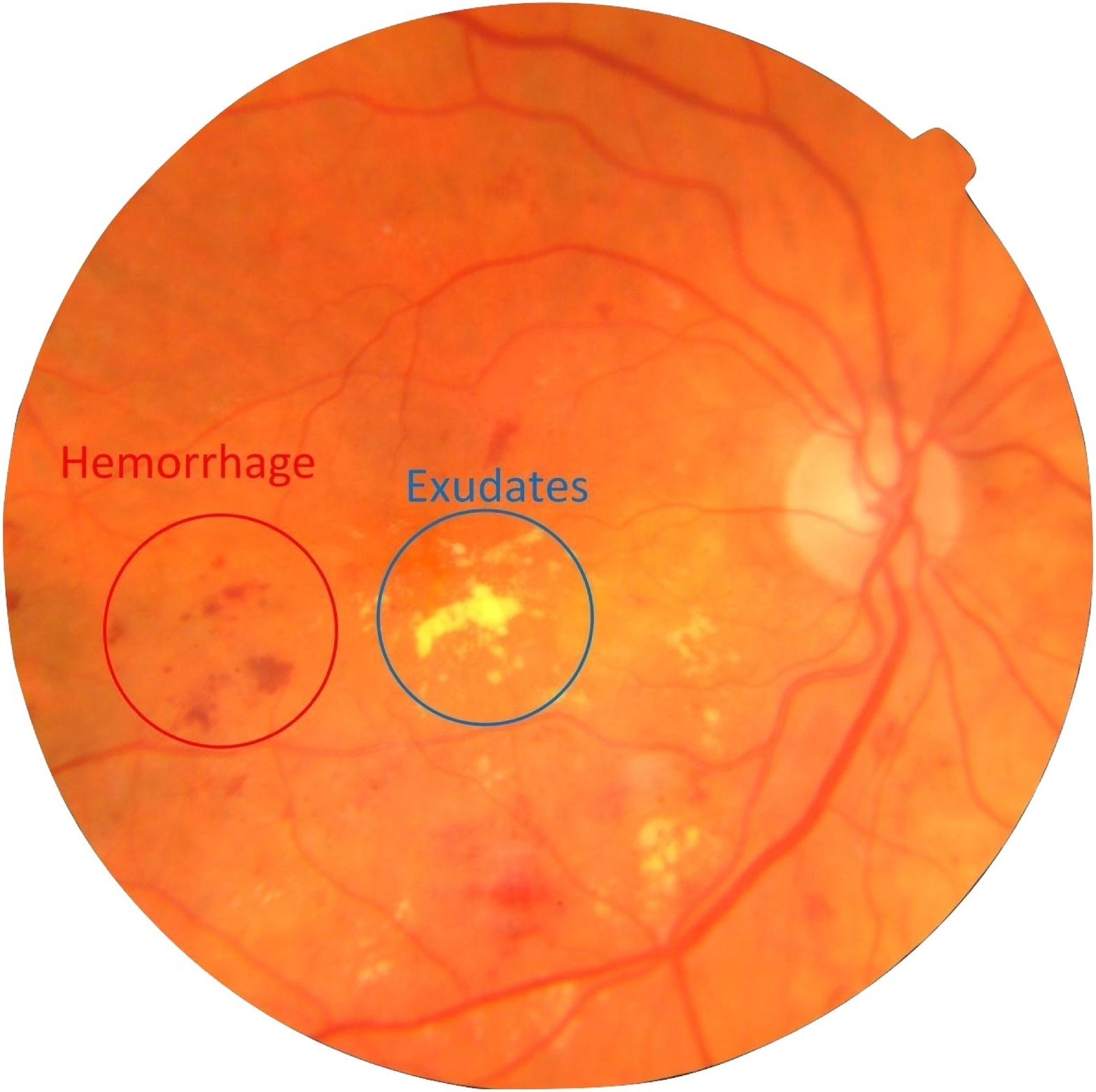}}
\caption{The figure showing (a) Normal image (b) Anatomical structure in retina (c)~Retinal image with exudates (d) Exudates in blue circle and haemorrhage in red circle.}
\label{exudates}
\end{figure}

Detailed examination and change analysis of retinal images helps in diagnosing various retinal eyes diseases \cite{khan2020shallow, khan2022t}. These diseases include glaucoma \cite{tabassum2020cded,rehman2019multi,imtiaz2021screening}, chronic systematic hypoxemia cataract \cite{Traustason2011}, diabetic retinopathy \cite{khan2016automatic,khan2020exploiting,khan2020semantically}, age-related mascular degeneration (AMD) \cite{Weeks1897} and retinal vascular occlusions \cite{fujio1994}. The progression, identification, and diagnosis of the disease at its earlier stage can save vision loss, especially in the case of mascular edema and diabetic retinopathy, leading to cost saving profitable treatment options. Exudates are lipoprotein deposits and lipids that appear near the retinal vessels inside the retina. They are evolved in the early stages of diabetic retinopathy and can be displayed as yellow areas that can be as large as a few pixels to an optical disc as shown in Figure \ref{exudates}. The precise computer-based automatic segmentation of the retinal symptoms (exudates) is critical for early detection and treatment of retinal-diseases that threaten vision. Therefore, the role of exudates segmentation to implement computerized retinal screening option is unrivaled.

To perform disease assessment from retinal images, one option is to manually divide images, a task carried out by trained optometrists or ophthalmologists \cite{khan2021rc,khan2022width}. It is a tedious and time-consuming task that requires extensive skills. In contrast, computerized segmentation algorithms can be used for large public screening with increased precision and reduced workload \cite{khan2021residual,naveed2021towards,khan2021leveraging}. However, automating exudates segmentation using computerized algorithms is quite challenging.

In this paper, we propose a supervised segmentation approach for automated exudates segmentation. The proposed supervised segmentation approach for exudates localization is independent of other anatomical information, variation in size and shape of the exudates. To improve the performance of network architecture, a suitable image augmentation technique is used to enhance the ability of the network. The proposed method employs the textural and pixel characteristics of regions for deep neural network classifier from negative and positive regional image samples. The network can robustly segment and localize exudates with high accuracy, which makes it suitable for diabetic retinopathy screening.

To summarize, in this paper, we make the following contributions:

\begin{enumerate}
 \item The exudates segmentation problem is modeled as a pixel-based classification problem through a semantic segmentation architecture.
  \item Retinal images are characterized by textural as well as pixel properties in a semantic-based classification framework that makes it robust in multicolored challenges in exudates localization.
  \item A comparison of different architectures is presented for exudates localization and segmentation in retinal images. Our supervised method illustrates that the improved segmentation results are superior to other segmentation based methods and unsupervised techniques.
\end{enumerate}

The structure of this paper is as follows. Detailed related work is described in section 2. Semantic segmentation is explained in Section 3. The proposed methodology is presented in Section 4. Section 5 describes the databases that are used as benchmark.  Section 6 presents experimental analysis followed by the conclusion in Section 7.

\section{\bf \textsc{Literature review}}
Computerized exudates segmentation and localization is one of the most challenging task and critical problem in retinal image analysis. The main challenges include variable physical characteristics of exudates (such as structure, size, and vessel structure) and various retinal pathology. Many OD, vessel and eye leakages detection protocols have been explored for 2D digital retinal images. The latest state-of-the-art on exudates segmentation and localization can be divided into two types: supervised and unsupervised methods. The segmentation and detection methods for exudates can be further divided into four main types:

\begin{enumerate}
  \item Mathematical morphology based
  \item Region growing
  \item Threshold-based
  \item Machine learning algorithm
\end{enumerate}
where all of these are unsupervised except machine learning algorithm which is supervised.

The authors in \cite{fraz2017multiscale}, explore methods based on mathematical morphological processes. In these methods, the primary structure (OD, blood vessels) is typically first segmented and removed from the image to reduce interference with exudates. Walter et al. \cite{walter2002contribution} applied morphological closure to eliminate retinal vessel, followed by threshold and local deviation to segment and localize the exudates. These techniques applied morphological rearrangements to determine the location of the exudate boundary. Morphological-based mathematical methods use many calculation operators with a variety of structural elements. Zhang et al. \cite{zhang2014exudate} used the morphological process to create applicant regions. The random forest and contextual features are then used to determine the actual positive pixels. In \cite{harangi2012automatic}, Hajdu and Harangi introduced  combination of morphological operators and active contour models. Mehdi et al. \cite{eadgahi2012localization} used morphological processes to reveal exudate conditions. In the pre-processing step, the OD and retinal vessels are eliminated by the help of morphological operators such as a bottom hat, top hat and reconstruction operators are used for exudate segmentation.

Region growing is a simple way to classify images by region. It is also classified as pixel-based image segmentation because it includes the choice of the initial seed point. This method checks the pixel next to the primary seed point and determines if pixels should be added to the region. This process is repeated in the same way as a general clustering algorithm \cite{chudzik2018exudates}. The method of region growing has also been explored to be useful for spontaneous exudates segmentation with a combination of artificial neural networks \cite{usher2004automated}. Chutatap and  Li.\cite{li2004automated} used the technique  of  edge detection with a combination of region growing method.

 Threshold-based methods exploit the variation in color intensity between different image regions. These methods depends on the global image grey-level or the local image grey-level \cite{phillips1993automated,wisaeng2015automatic}.  Pereira et al. \cite{pereira2015exudate} combined threshold method with the optimizer of the ant colony to segment exudates. Threshold methods were used to identify exudates candidates region, while ant colony optimizers were used to improve the edges of exudates. Garcia et al. \cite{garcia2009detection} proposed a collective combination of global and adaptive threshold approaches to differentiate the applicant region of exudates. Subsequently, based on the radial design basic function, the lesions are classified based on features that are primarily derived from the color and shape of the lesion.

Finally, many machine learning methods have been introduced in the literature such as support vector machine \cite{fleming2007automated,giancardo2012exudate},  random forest \cite{zhang2014exudate}, discriminant classifier \cite{sanchez2008novel,niemeijer2007automated} and Na\"ive Bayes classifier. These methods commonly start with the normalization of the image, followed by OD localization and segmentation. Next, a set of candidate regions i.e., structures same as exudates are identified and then the classification exudates region and features computation of each applicant region is defined. In literature, most methods used a feature vector for a single pixel or cluster of pixels and then a machine learning method is used to classify these feature vectors into exudates and no exudates. These features are commonly based on size, color, brightness, edges, textures information and other variations of pixel cluster. A concise summary of previous related methods is given in Table.~\ref{review}.

\begin{table*}[!t]
\caption{Summary of previous algorithm used for exudates with different approaches in retinal image processing}
\resizebox{\columnwidth}{!}{%
        \begin{tabular}{ccp{17em}p{7.111em}r}
    \hline
    \multicolumn{1}{p{1.678em}}{\textbf{Sr. No.}} & \textbf{Author} & \multicolumn{1}{c}{\textbf{Method}} & \textbf{Database used } & \multicolumn{1}{p{7.15em}}{\textbf{Classification Technique}} \bigstrut\\
    \hline
    \multirow{3}[2]{*}{1} & \multicolumn{1}{c}{\multirow{3}[2]{*}{\cite{zhang2014exudate}}} & A segmentation based on mathematical & E-Ophtha  & \multicolumn{1}{p{7em}}{Random forest} \bigstrut[t]\\
          &       & morphology using contextual features  & ----- & \multicolumn{1}{p{7em}}{classiﬁer } \\
          &       &and classical features. & ----- &  \bigstrut[b]\\
    \hline
    \multirow{3}[2]{*}{2} & \multirow{3}[2]{*}{ \cite{prentavsic2016detection}} & A convolutional neural network & DIARETDB1  & \multicolumn{1}{p{7em}}{Deep neural} \bigstrut[t]\\
          &       &combine with landmark detectors. & ----- & \multicolumn{1}{p{7em}}{networks} \\
          &       & \multicolumn{1}{r}{} & ----- &  \bigstrut[b]\\
    \hline
    \multirow{3}[2]{*}{3} & \multirow{3}[2]{*}{\cite{pereira2015exudate}} & A unsupervised method based on  & HEI-MED & \multicolumn{1}{p{7em}}{Ant colony } \bigstrut[t]\\
          &       & algorithm the ant colony  & ----- & \multicolumn{1}{p{7em}}{optimization} \\
          &       & optimization.  & ----- &  \bigstrut[b]\\
    \hline
    \multirow{3}[2]{*}{4} & \multirow{3}[2]{*}{\cite{feng2017deep}} & FCNs used introducing long and  & DRIONS-DB & \multicolumn{1}{p{7em}}{Fully } \bigstrut[t]\\
          &       & short skip  connections with  & Personal & \multicolumn{1}{p{7em}}{convolution } \\
          &       & class-balancing loss.   & ----- & \multicolumn{1}{p{7em}}{neural network} \bigstrut[b]\\
    \hline
    \multirow{2}[1]{*}{5} & \multirow{2}[1]{*}{\cite{almotiri2018multi}} & A hybrid segmentation algorithm  & DIARETDB1 & \multicolumn{1}{p{7em}}{Fuzzy C-means} \bigstrut[t]\\
          &       & with mathematical morphology  & ----- &  \\
          &       & and adaptive fuzzy thresholding  & \multicolumn{1}{c}{} &  \bigstrut[b]\\
    \hline
    \multirow{3}[2]{*}{6} & \multirow{3}[2]{*}{\cite{fraz2017multiscale}} & An ensemble classifier of bootstrapped  & DIARETDB1 & \multicolumn{1}{p{7em}}{ Ensemble } \bigstrut[t]\\
          &       & decision trees combines with Gabor & E-Ophtha  & \multicolumn{1}{p{7em}}{classiﬁer} \\
          &       &  filter and morphological reconstruction.  & HEI-MED &  \bigstrut[b]\\
    \hline
    \end{tabular}%

    }
  \label{review}%
\end{table*}%

Convolutional neural networks (CNN) exhibit excellent performance in a variety of applications including medical image processing. Saven et al. \cite{prentavsic2016detection} designed a 10-layer neural network to detect exudates. In \cite{feng2017deep}, Feng et al. proposed the application of Fully Adaptive Neural Network in the classification of exudates and OD. Fujita et al. \cite{tan2017automated} utilized a single neural network to detect different symptoms such as hemorrhages, micro-aneurysms and exudates simultaneously. CNN-based techniques are considered promising for the analysis of fundus images. However, compared to modern complex methods, the performance of CNN-based methods can be improved further, especially for the DiaReTDB1 database \cite{feng2017deep}, as sensitivity obtained is 81.35\%. The poor performance of the CNN method comes in part from limited training data and class unbalanced data. Unbalanced data poses significant challenges for many standard classifiers, most of which have been optimized to lessen global error rates without regard to data distribution. One of the direct ways to alleviate the problem of class imbalance is to raise the level of the minority or reduce the majority class samples. In this regard, SMOTE algorithm \cite{chawla2002smote} is considered to be an effective technique to tackle the class imbalance problem.

A supervised technique for exudates segmentation based on pixel-wise features and transfer learning is proposed recently in \cite{chudzik2018exudate}. This method trains an FCCN architecture with inception module and loss function of the Dice coefficient for exudates detection. The exudates are automatically segmented by this technique as compared to the previous methods, which rely on detection and removal of healthy retinal structure first before exudate detection can be initiated. Although the method proposed by Chudzik et al. \cite{chudzik2018exudate} does not rely on removal of healthy anatomical structures, it involves a few pre-processing steps before the patches can be employed for pixel-wise classification. Recent methods \cite{feng2017deep,chudzik2018exudate,zheng2018detection} of deep neural networks have been studied for segmentation of exudate with special properties in cross-training and separate-training assessments. Due to the high precision and generalization of invisible data, neural network-based methods are promising for real-time scanning and diagnostics.

The high precision in neural network-based methods comes at the cost of a large amount of training data and higher computational time. The challenge is to create a variation in augmented data, when dataset is small, that captures the basic distribution of highly varying retinal images. The high degree of diversity between datasets makes this problem very difficult with only a limited number of training images. The computational time requirements of deep neural network training is another factor limiting their application in diagnostic and screening applications. While large clinical centers can contain advanced computing resources, not all scanning facilities have the capability of powerful computing. In addition, large training times make synchronized training of new patient data impractical.

The proposed method is advantageous in the case of training data scarcity. Since the proposed method depends on the region, a limited number of training images is sufficient to produce a reasonable number of training cases. Low training time and minimum computational ability make the proposed method a suitable candidate for online updating and training of new patient record. Another noteworthy feature of the proposed method is that unlike previous methods, it uses a number of modalities and cues to obtain predictions on the pixel-wise region. This can essentially distinguish exudates from other symptoms in retinal structures and provide a robust image for pathology.

\section{\bf \textsc{Semantic segmentation}}
Semantic segmentation is the process of linking each pixel image to a class, resulting in an image that is divided by class category. It has been broadly used in a variety of applications, including autonomous driving, medical image analysis, scene understanding, and industrial inspection. Although automatic learning algorithms such as boosting \cite{shotton2008semantic} and random forest \cite{ladicky2010and}, have traditionally been used for semantic segmentation, however, convolutional neural networks (CNN's) have gained significant advances in this area.

An overall semantic segmentation architecture can be widely considered as a coding network followed by a decoder network:
\begin{itemize}
\item The encoder is usually a pre-trained classification network such as VGG / ResNet /Unet followed by a decoder network.
\item The function of the decoder is to visualize semantically the resolution characteristics of the encoder (lower resolution) to the pixel state (higher resolution) to obtain a dense rating.
\end{itemize}

\begin{figure*}[!t]
\centering
\includegraphics[width=4.5in]{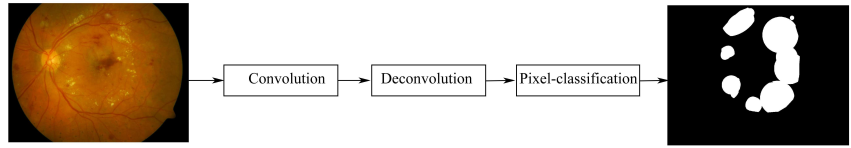}
\caption{General structure of encoder-decoder methods for semantic segmentation.}
\label{SemanticSegmentation}
\end{figure*}

Unlike the categorization where the deep network is the only important thing, semantic segmentation requires not only pixel level discrimination but also the mechanism by which the projection at different stages of the encoder learns the separation characteristics of the pixel spacing. Different approaches used different mechanisms as a fragment of the decoding mechanism. The general structure of the encoder-decoder methods for semantic segmentation is presented in Figure~\ref{SemanticSegmentation}.

\section{\bf \textsc{Methodology}}
This section presents semantic segmentation, standardized RTC-net architecture to detect exudates and demonstrated  systematic data augmentation to generate training sample of images and briefly introduction of benchmark databases. The proposed method training uses E-ophthta and also further evaluated on HEI-MED and E-ophtha and DiaReTDB1 dataset.
\begin{figure}[!b]
\centering
\includegraphics[width=2in]{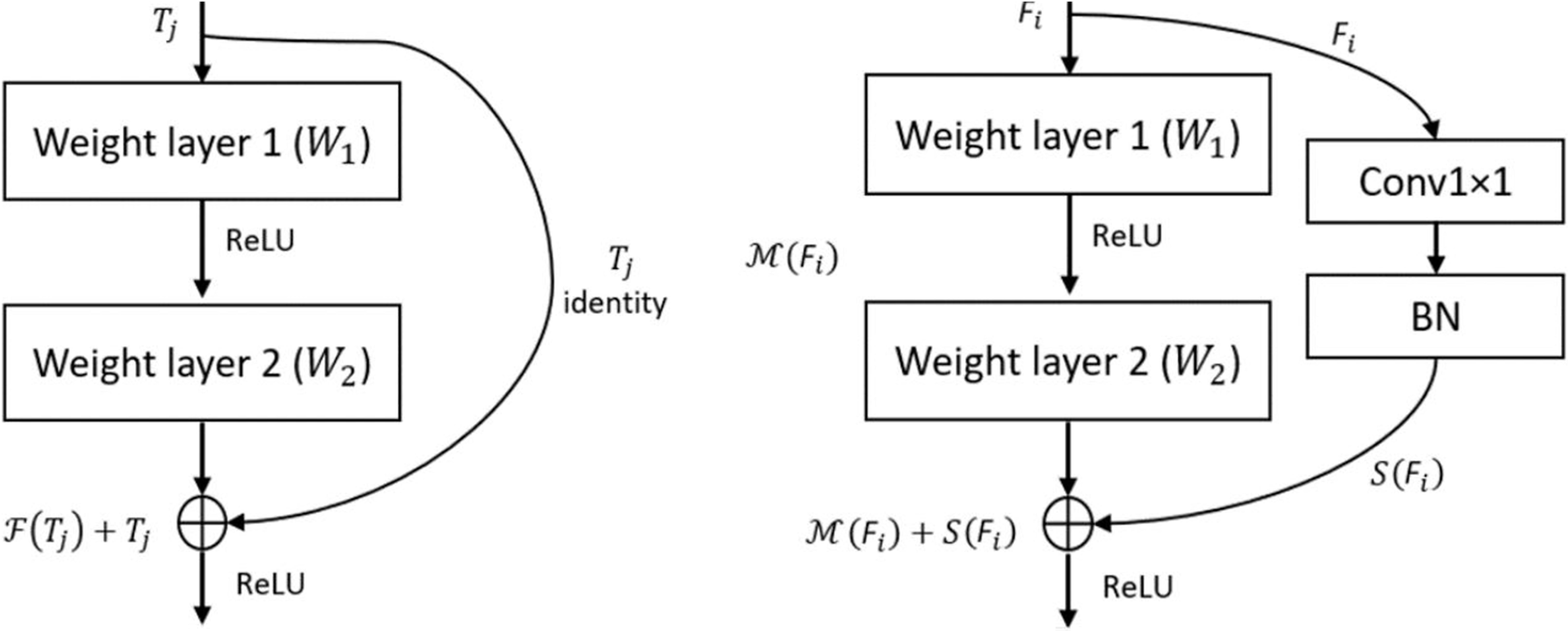}
\caption{Sample of skipping connection in RTC network architecture.}
\label{RTC}
\end{figure}
%\begin{figure*}[!t]
%\centering
%\includegraphics[width=6.9in,height=3in]{B1.eps}
%\caption{Residual Encoder and decoder block}
%\label{B1}
%\end{figure*}
\begin{figure*}[!b]
\centering
\includegraphics[width=4.5in]{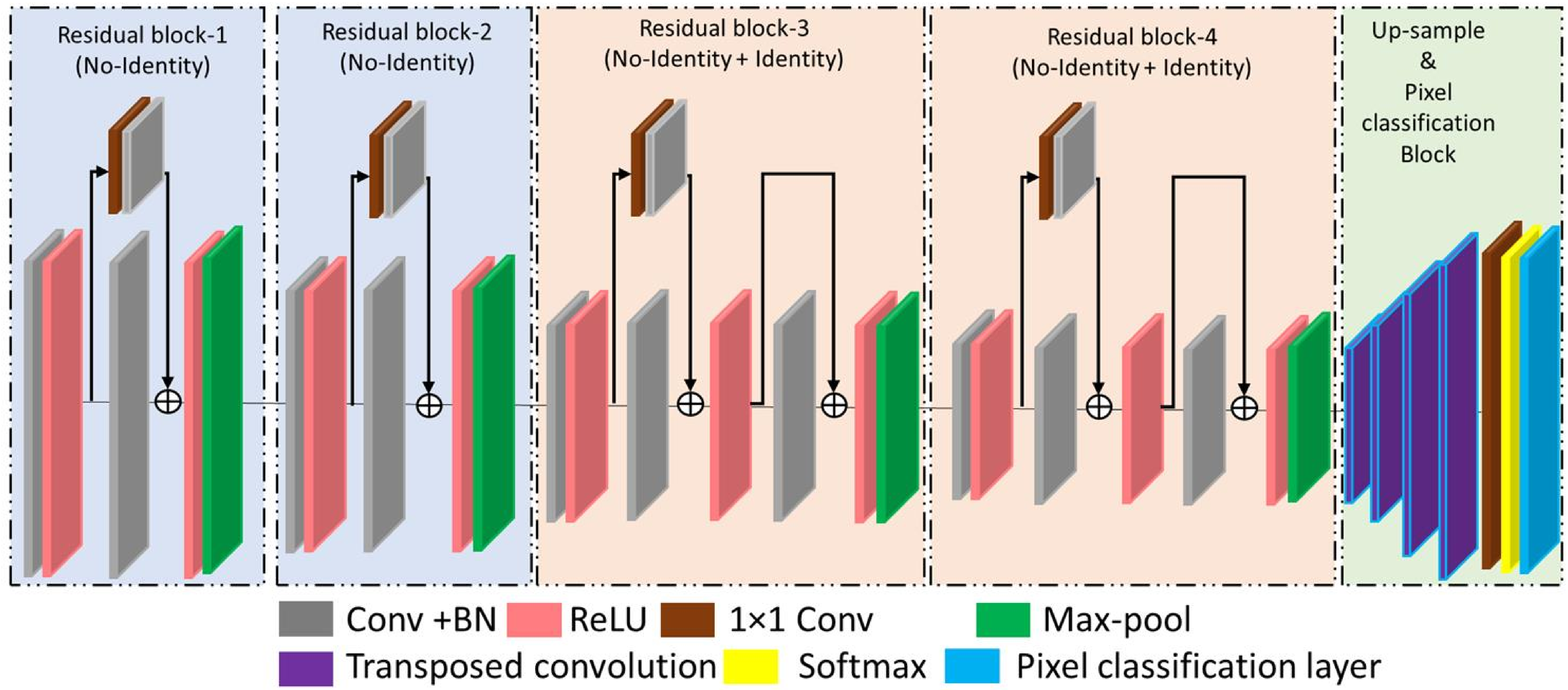}
\caption{Residual Encoder and decoder block}
\label{B2}
\end{figure*}

\subsection{\bf \textsc{Data augmentation}}
There are two options for maximizing data through augmentation. The first one (type-1) is termed as offline augmentation. It is preferred in a relatively small database because the size of the database is eventually increased by a factor equal to the number of changes made (by flipping all the images, the size of the database becomes 2x). The second one (type-2) is called online data augmentation, or instant incrementing. For larger databases, this method is mostly preferred to keep size limited. In this method, performing the transformation in mini-batches is preferred. Some computer-based learning algorithms support online augmentation process that can be accelerated on the graphics processing unit (GPU).

In this paper, more than two thousand fundus images are obtained from less than one hundred images after cropping, resizing, magnifying, translation and flipping each image in the database making it possible to generate synthetic data having multiple copies of previous data with different contrast variation and illumination change. Additionally, augmented data is divided into testing and training images folder for the training process.

\subsection{ \bf \textsc {Residual convolution network}}
\label{Res}
Considering the deep learning based semantic segmentation there are two famous approaches. The first one is Fully Convolutional architectures FCN, using Alexnet, VGG-16 and GoogleNet (FCN-Alexnet, FCN-VGG, FCN-GoogleNet) \cite{long2015fully}. The second approach is SegNet encoder decoder-based networks. The networks that are based on FCN consumes more parameters which make the method computationally inefficient, whereas the architectures based on encoder decoder are complex in nature. The reason for this is because similar decoder should be designed in order to upsample the image back to its original size. The same number of layers in decoder increases the number of layers as well as the depth of the network.

This study is aims to reduce the network depth and parameters, so the semantic segmentation of exudates is obtained by just $10$ convolutional layers as shown in Table.~\ref{layers}. There are the few advantages that can be deduced by the proposed network.

\begin{enumerate}
 \item The usual semantic segmentation involves encoder decoder pattern like in SegNet $26$ convolutional layers ($13$ encoder $13$ decoder), in which the decoder is a mirror copy of the encoder to up-sample the image back to its original size. This mirror copy of the encoder involves same number of trainable parameters which make the network computationally inefficient. RTC-Net350 does not use the mirror copy decoder but it uses few transposed convolution layers to up-sample the image with few parameters. ($11.3$ Million parameters total).
 \item Conventional networks are VGG-Net type series networks which involve many convolutions and the gradient vanishes in these convolutions continuously, which results in compromised accuracy of the network. The RTC-Net involves the residual connectivity which empowers the feature by importing the features from previous convolution layer by a residual skip connection in the form of identity and non-identity mapping, as shown in Figure~\ref{RTC}. The feature S is the empowered feature after addition layer which removes the effect of gradient loss during the sampling process (see Figure~\ref{RTC}).
\item RTC-Net only uses $10$ convolutional layers of $3\times3$ and four transposed convolutions for up-sampling, where for mapping four non-identity skip connections are created using $1\times1$ convolutions in skip paths.
\item There are two types of mapping used in our proposed RTC-Net, one is non-identity mapping used in ResNet \cite{he2016deep}, which is used to add layers with a different number of $370$ channels. Where other is identity mapping by which similar channel size features can be added.
\item RTC-Net reduces the networks complexity and up-samples the image using un-pooling layers (unpooling layers are also used in SegNet).
\end{enumerate}
The overall segmentation process is carried out in three blocks Residual convolutional block, Up-sampling block and Pixel classification block as shown in Figure.~\ref{B2} and Table~\ref{layers}.

\subsubsection{\bf \textsc{Residual block}}
In proposed network, the input image is provided to the first residual block as shown in  Figure.~\ref{B2}. There are total 4 residual convolutional blocks with total 10 $3\times3$ convolutional layers. The first two blocks are with $2$ convolutional layers which relate to non-identity residual connectivity to compensate the feature loss. The residual block $3$ and $4$ have $3$ convolutional layers which use both identity and non-identity connectivity between the layers. The 1$\times$1 convolution are used to match the number of channels for non-identity mapping. During the convolution process in all the residual blocks, the image sampling and the feature empowerment is carried out at the same time. The final feature map size after the pool-4 (as shown in Table~\ref{layers} ) is $28 \times 32\times 512$ which is a rich feature with better edge information. This empowered feature is used to up-sample in the next block.

\subsubsection{\bf \textsc{Up-sampling block}}
In order to up-sample the image to its original size the four transposed convolutions are used one after another. The $28 \times 32\times512$ feature from the previous block is then up-sampled back to $448 \times 512 \times 64$ as shown in Table.~\ref{upsample}. The transposed convolution ensures appropriate upsampling is applied as the weights are learnable.

\subsubsection{\bf \textsc{Pixel-Classification block}}
After the up-sampling of the image in up-sampling block $448\times 512$ feature with depth of $512$ channels are provided to a $1\times1$ convolution which is used to control the number of channels according to the number of classes. In our case, the number of classes are two so the number of filters of $1\times1$ convolution (Con-class as shown in Table.~\ref{upsample}) are set at 2. Then in the final layers the softmax loss is used for the classification purposes. Finally, the pixel classification layers provide each pixel classification according to the loss function. The output of the network is a binary mask with two classes.
\begin{table*}[!t]
\centering
\caption{The architectural level and parametric details of the proposed method. }
\resizebox{\columnwidth}{!}{%
        \begin{tabular}{rrrrr}
    \hline
    \multicolumn{1}{p{7.93em}}{\textbf{Block}} & \multicolumn{1}{p{2.215em}}{\textbf{Filter}} & \multicolumn{1}{p{13.855em}}{\textbf{Name (Size)}} & \multicolumn{1}{p{7.5em}}{\textbf{Activation Size}} & \multicolumn{1}{p{8.645em}}{\textbf{No. of Parameters}} \bigstrut\\
    \hline
    \multicolumn{1}{p{7.93em}}{Input} &       & \multicolumn{1}{p{13.855em}}{Input Image} & \multicolumn{1}{p{7.5em}}{448 × 512 × 3} & \multicolumn{1}{p{8.645em}}{ } \bigstrut\\
    \hline
    \multicolumn{1}{p{7.93em}}{Residual block-1} & \multicolumn{1}{c}{64} & \multicolumn{1}{p{13.855em}}{Con-1a (3 × 3 × 3)} &       & \multicolumn{1}{c}{1792} \bigstrut\\
\cline{2-5}    \multicolumn{1}{p{7.93em}}{(Non-Identity)} & \multicolumn{1}{p{2.215em}}{ } & \multicolumn{1}{p{13.855em}}{Relu-1a} &       & \multicolumn{1}{p{8.645em}}{ } \bigstrut\\
\cline{2-5}          & \multicolumn{1}{c}{64} & \multicolumn{1}{p{13.855em}}{Con-1b (3 × 3 × 64)} & \multicolumn{1}{p{7.5em}}{448 × 512 × 64} & \multicolumn{1}{c}{36928} \bigstrut\\
\cline{2-5}          & \multicolumn{1}{c}{64} & \multicolumn{1}{p{13.855em}}{Skip-1* (1 × 1 × 64)} &       & \multicolumn{1}{c}{4160} \bigstrut\\
\cline{2-5}          & \multicolumn{1}{p{2.215em}}{  } & \multicolumn{1}{p{13.855em}}{Add-1 (Skip-1* + Con-1b)} &       & \multicolumn{1}{p{8.645em}}{ } \bigstrut\\
\cline{2-5}          &       & \multicolumn{1}{p{13.855em}}{Relu-1b} &       &  \bigstrut\\
\cline{2-5}          & \multicolumn{1}{p{2.215em}}{ } & \multicolumn{1}{p{13.855em}}{Pool-1 (2 × 2)} & \multicolumn{1}{p{7.5em}}{224 × 256 × 64} & \multicolumn{1}{p{8.645em}}{ } \bigstrut\\
    \hline
    \multicolumn{1}{p{7.93em}}{Residual block-2} & \multicolumn{1}{c}{128} & \multicolumn{1}{p{13.855em}}{Con-2a (3 × 3 × 64)} &       & \multicolumn{1}{c}{73856} \bigstrut\\
\cline{2-5}    \multicolumn{1}{p{7.93em}}{(Non-Identity)} & \multicolumn{1}{p{2.215em}}{ } & \multicolumn{1}{p{13.855em}}{Relu-2a} &       & \multicolumn{1}{p{8.645em}}{ } \bigstrut\\
\cline{2-5}          & \multicolumn{1}{c}{128} & \multicolumn{1}{p{13.855em}}{Con-2b (3 × 3 × 128)} & \multicolumn{1}{p{7.5em}}{224 × 256 × 128} & \multicolumn{1}{c}{147584} \bigstrut\\
\cline{2-5}          & \multicolumn{1}{c}{128} & \multicolumn{1}{p{13.855em}}{Skip-2* (1 × 1 × 128)} &       & \multicolumn{1}{c}{16512} \bigstrut\\
\cline{2-5}          & \multicolumn{1}{p{2.215em}}{ } & \multicolumn{1}{p{13.855em}}{Add-2 (Skip-2* + Con-2b)=Relu-2b} &       & \multicolumn{1}{p{8.645em}}{ } \bigstrut\\
\cline{2-5}          & \multicolumn{1}{p{2.215em}}{ } & \multicolumn{1}{p{13.855em}}{Pool-2 (2 × 2)} & \multicolumn{1}{p{7.5em}}{112 × 128 × 128} & \multicolumn{1}{p{8.645em}}{ } \bigstrut\\
    \hline
    \multicolumn{1}{p{7.93em}}{Residual block-3} & \multicolumn{1}{c}{256} & \multicolumn{1}{p{13.855em}}{Con-3a (3 × 3 × 128)} &       & \multicolumn{1}{c}{295168} \bigstrut\\
\cline{2-5}    \multicolumn{1}{p{7.93em}}{(Non-Identity} & \multicolumn{1}{p{2.215em}}{ } & \multicolumn{1}{p{13.855em}}{Relu-3a} &       & \multicolumn{1}{p{8.645em}}{ } \bigstrut\\
\cline{2-5}    \multicolumn{1}{p{7.93em}}{ + Identity)} & \multicolumn{1}{c}{256} & \multicolumn{1}{p{13.855em}}{Con-3b (3 × 3 × 256)} &       & \multicolumn{1}{c}{590080} \bigstrut\\
\cline{2-5}          & \multicolumn{1}{c}{256} & \multicolumn{1}{p{13.855em}}{Skip-3* (1 × 1 × 256)} &       & \multicolumn{1}{c}{65792} \bigstrut\\
\cline{2-5}          & \multicolumn{1}{p{2.215em}}{ } & \multicolumn{1}{p{13.855em}}{Add-3a (Skip-3* + Con-3b)} & \multicolumn{1}{p{7.5em}}{112 × 128 × 256} & \multicolumn{1}{p{8.645em}}{ } \bigstrut\\
\cline{2-5}          & \multicolumn{1}{p{2.215em}}{ } & \multicolumn{1}{p{13.855em}}{Relu-3b} &       &  \bigstrut\\
\cline{2-5}          & \multicolumn{1}{c}{256} & \multicolumn{1}{p{13.855em}}{Con-3c (3 × 3 × 256)} &       & \multicolumn{1}{c}{590080} \bigstrut\\
\cline{2-5}          & \multicolumn{1}{p{2.215em}}{ } & \multicolumn{1}{p{13.855em}}{Add-3b (Relu-3b + Con-3c)=Relu-3c} &       & \multicolumn{1}{p{8.645em}}{ } \bigstrut\\
\cline{2-5}          & \multicolumn{1}{p{2.215em}}{ } & \multicolumn{1}{p{13.855em}}{Pool-3 (2 × 2)} & \multicolumn{1}{p{7.5em}}{54 × 64× 256} &  \bigstrut\\
    \hline
    \multicolumn{1}{p{7.93em}}{Residual block-4} & \multicolumn{1}{c}{512} & \multicolumn{1}{p{13.855em}}{Con-4a (3 × 3 × 256)} &       & \multicolumn{1}{c}{1180160} \bigstrut\\
\cline{2-5}    \multicolumn{1}{p{7.93em}}{(Non-Identity } & \multicolumn{1}{p{2.215em}}{} & \multicolumn{1}{p{13.855em}}{Relu-4a} &       & \multicolumn{1}{p{8.645em}}{ } \bigstrut\\
\cline{2-5}    \multicolumn{1}{p{7.93em}}{+ Identity)} & \multicolumn{1}{c}{512} & \multicolumn{1}{p{13.855em}}{Con-4b (3 × 3 × 512)} &       & \multicolumn{1}{c}{2359808} \bigstrut\\
\cline{2-5}          & \multicolumn{1}{c}{512} & \multicolumn{1}{p{13.855em}}{Skip-4* (1 × 1 × 512)} &       & \multicolumn{1}{c}{262656} \bigstrut\\
\cline{2-5}          & \multicolumn{1}{p{2.215em}}{ } & \multicolumn{1}{p{13.855em}}{Add-4a (Skip-4* + Con-4b)} & \multicolumn{1}{p{7.5em}}{54 × 64× 512} & \multicolumn{1}{p{8.645em}}{ } \bigstrut\\
\cline{2-5}          & \multicolumn{1}{p{2.215em}}{ } & \multicolumn{1}{p{13.855em}}{Relu-4b} &       & \multicolumn{1}{p{8.645em}}{ } \bigstrut\\
\cline{2-5}          & \multicolumn{1}{c}{512} & \multicolumn{1}{p{13.855em}}{Con-4c (3 × 3 × 512)} &       & \multicolumn{1}{c}{2359808} \bigstrut\\
\cline{2-5}          & \multicolumn{1}{p{2.215em}}{ } & \multicolumn{1}{p{13.855em}}{Add-4b (Relu-4b + Con-4c)=Relu-4c} &       & \multicolumn{1}{p{8.645em}}{ } \bigstrut\\
\cline{2-5}          & \multicolumn{1}{p{2.215em}}{ } & \multicolumn{1}{p{13.855em}}{Pool-4 (2 × 2)} & \multicolumn{1}{p{7.5em}}{28 × 32× 512} & \multicolumn{1}{p{8.645em}}{ } \bigstrut\\
    \hline
          &       &       &       &  \bigstrut[t]\\
    \end{tabular}%

    }
  \label{layers}%
\end{table*}%
\subsection{\bf \textsc{Training and testing of the networks}}
\begin{figure*}[!b]
\centering
\subfigure[] {\label{cropping_a}\includegraphics[width=1in]{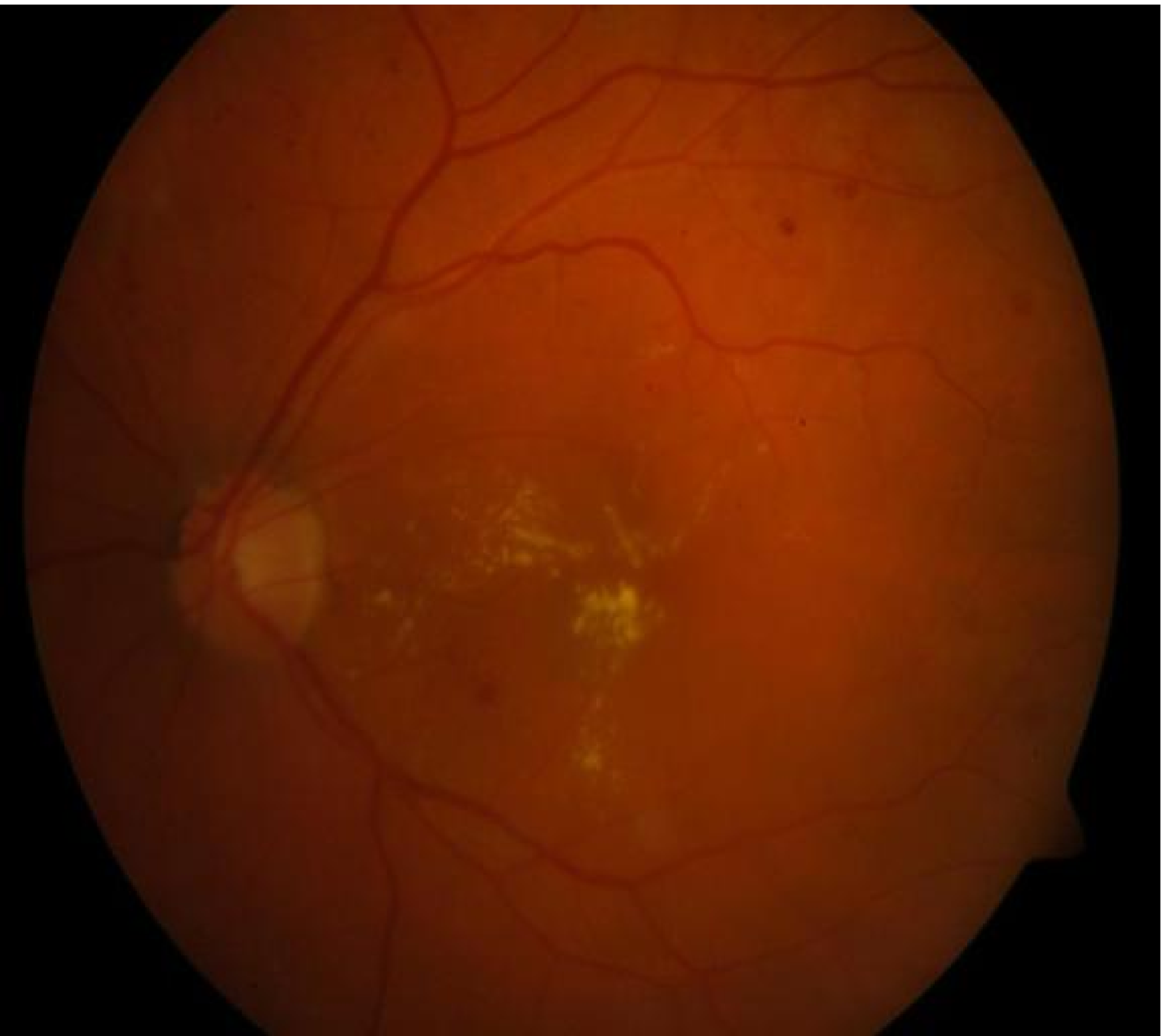}}
\subfigure[] {\label{cropping_a}\includegraphics[width=1in]{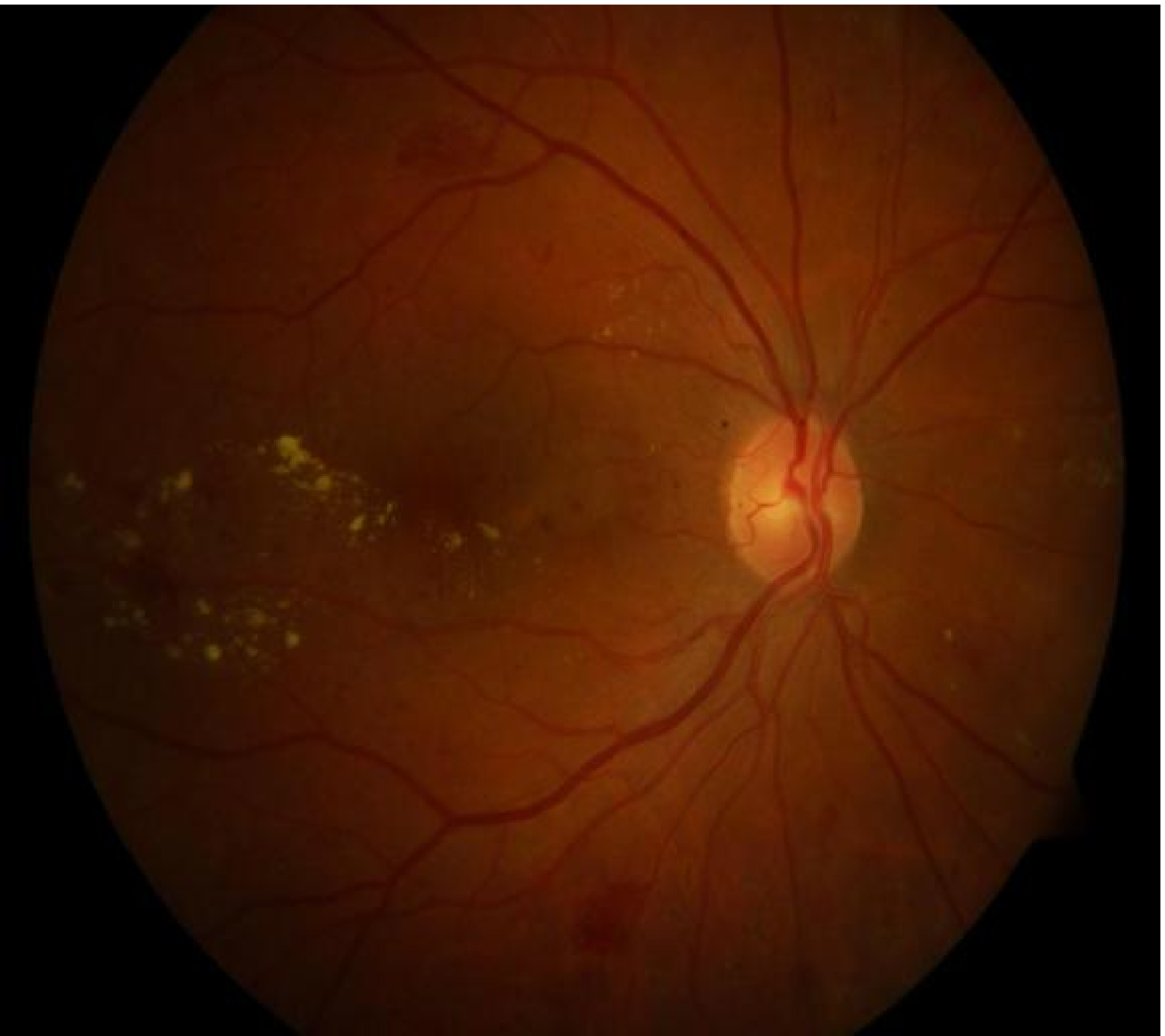}}
\subfigure[] {\label{cropping_a}\includegraphics[width=1in]{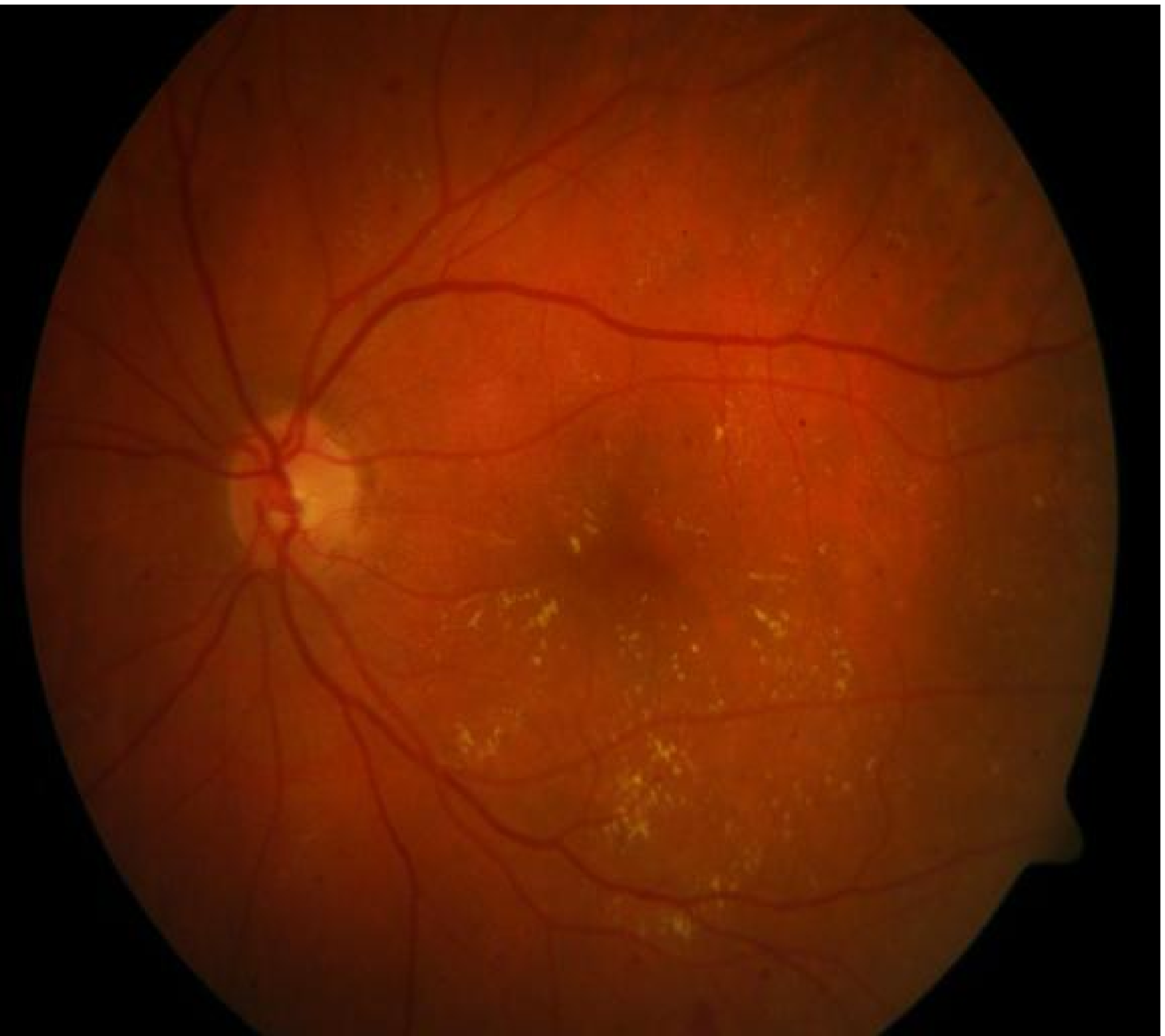}}
\subfigure[] {\label{cropping_a}\includegraphics[width=1in]{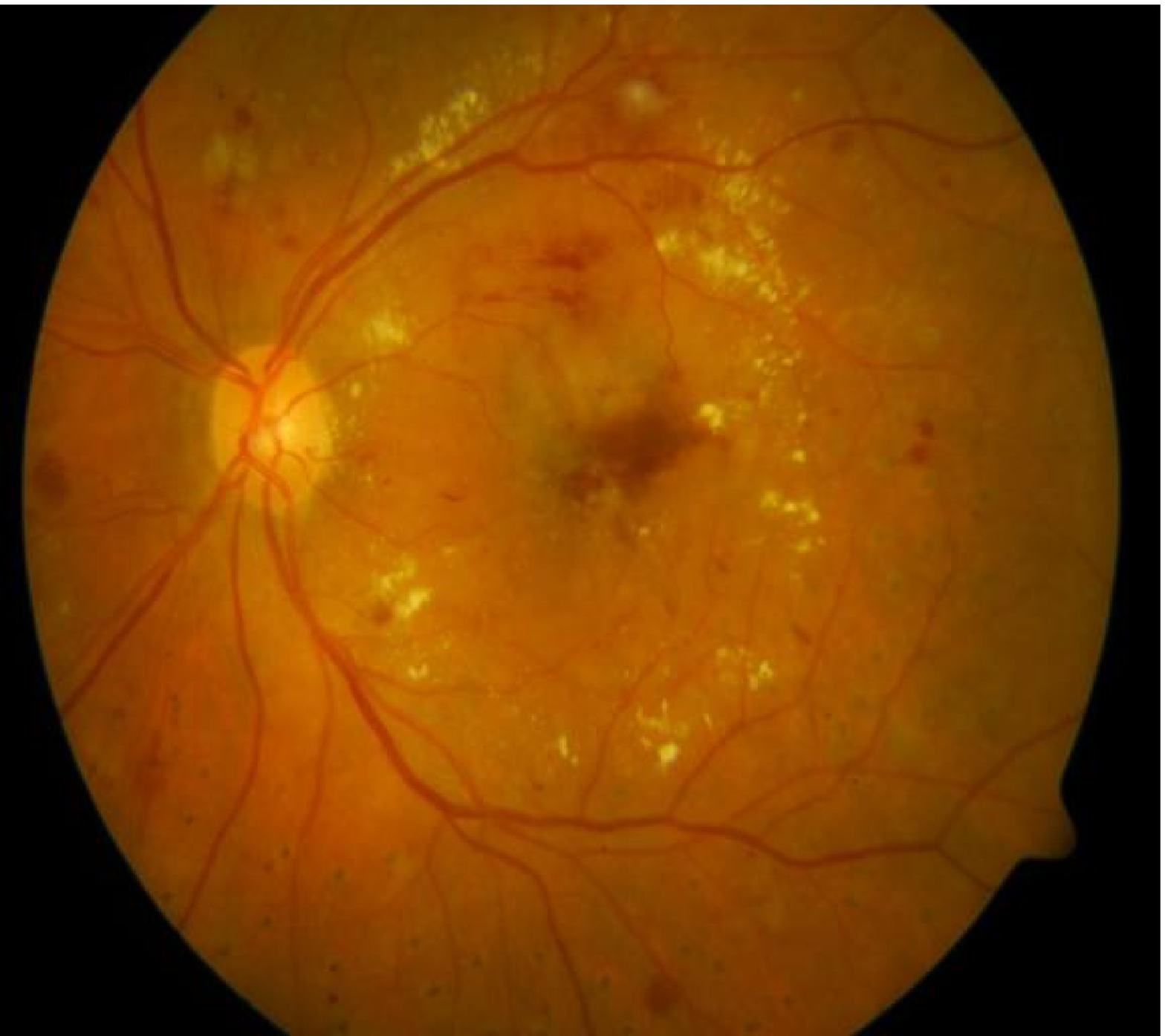}}

\subfigure[] {\label{cropping_a}\includegraphics[width=1in]{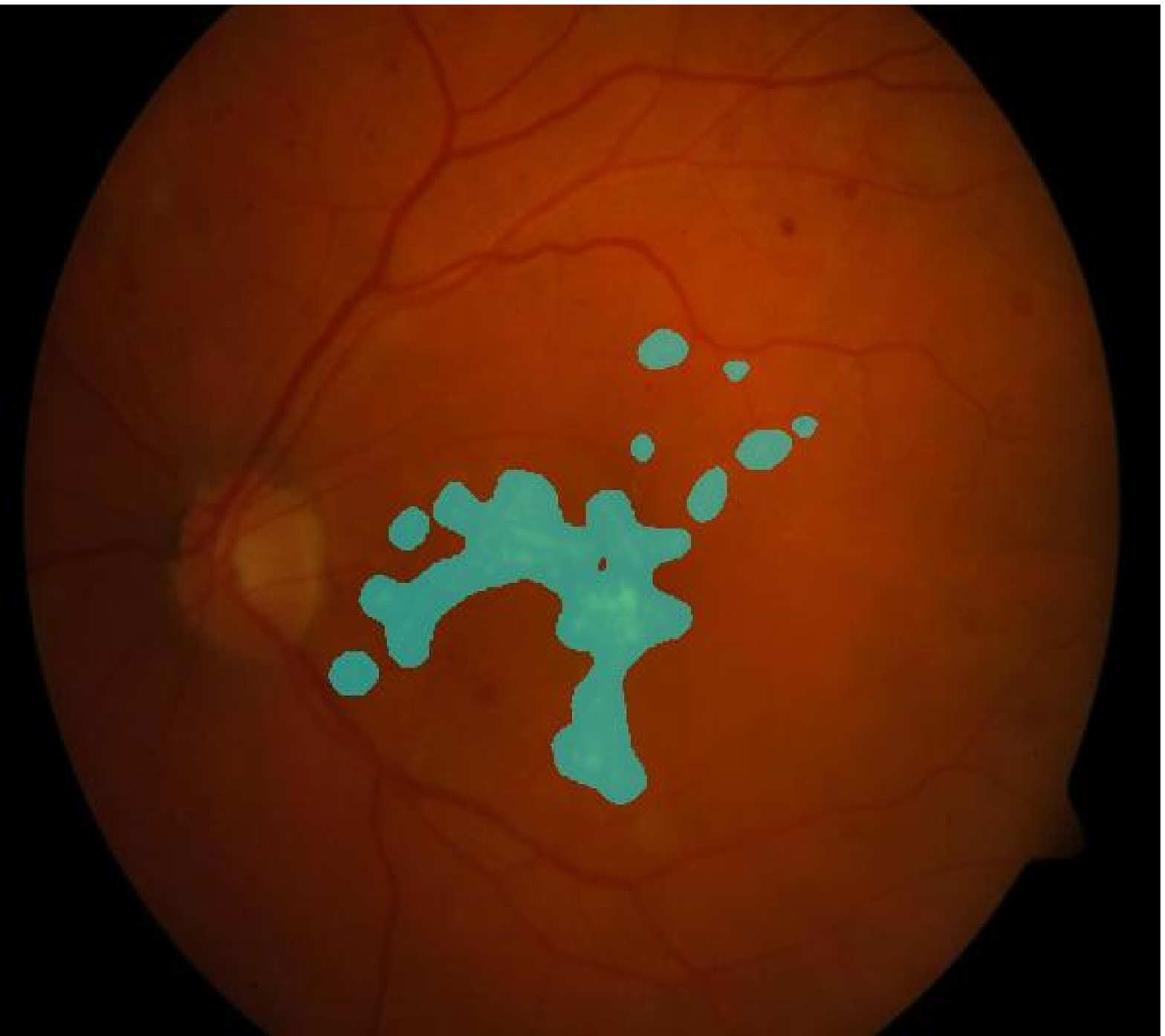}}
\subfigure[] {\label{cropping_a}\includegraphics[width=1in]{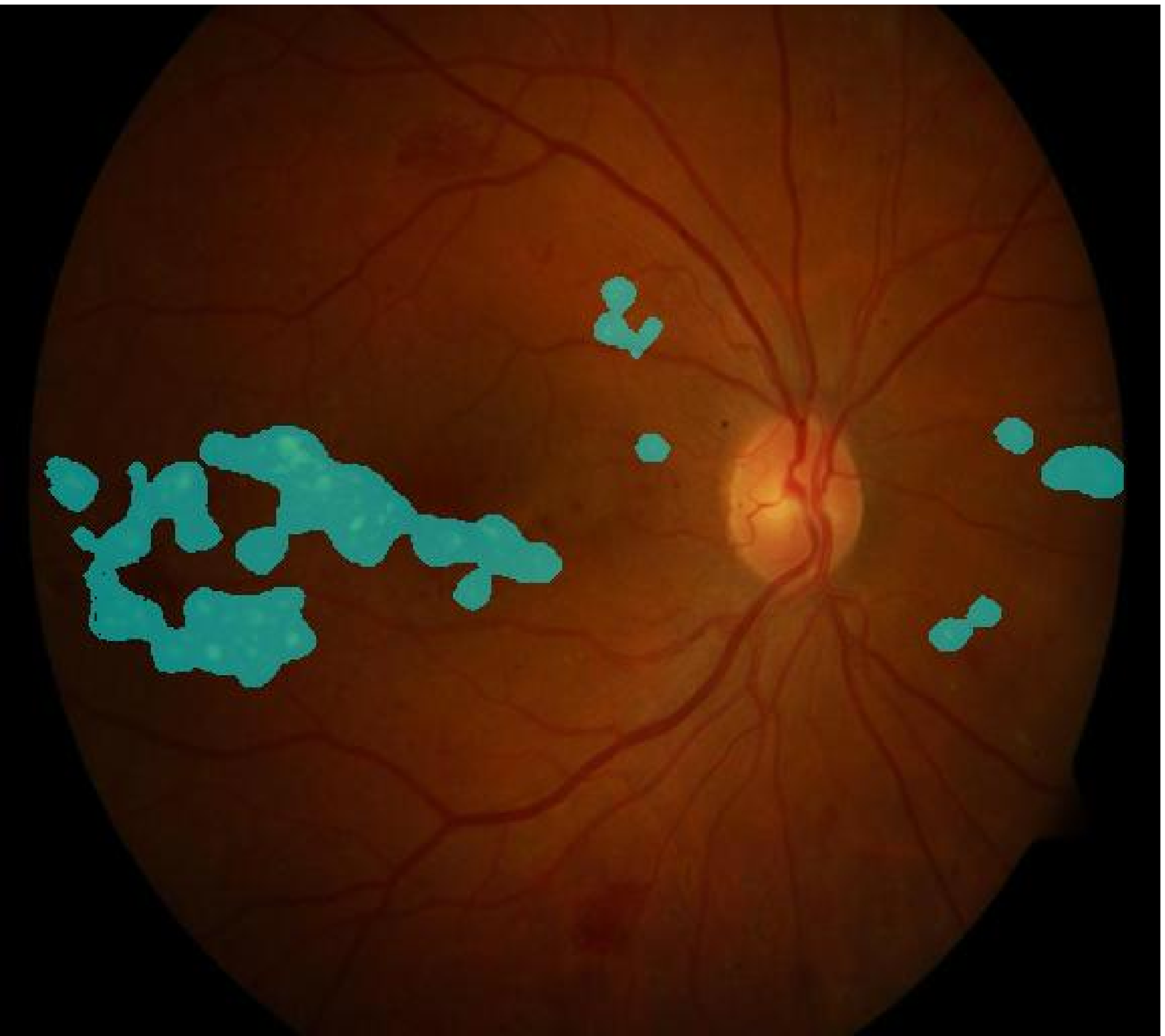}}
\subfigure[] {\label{cropping_a}\includegraphics[width=1in]{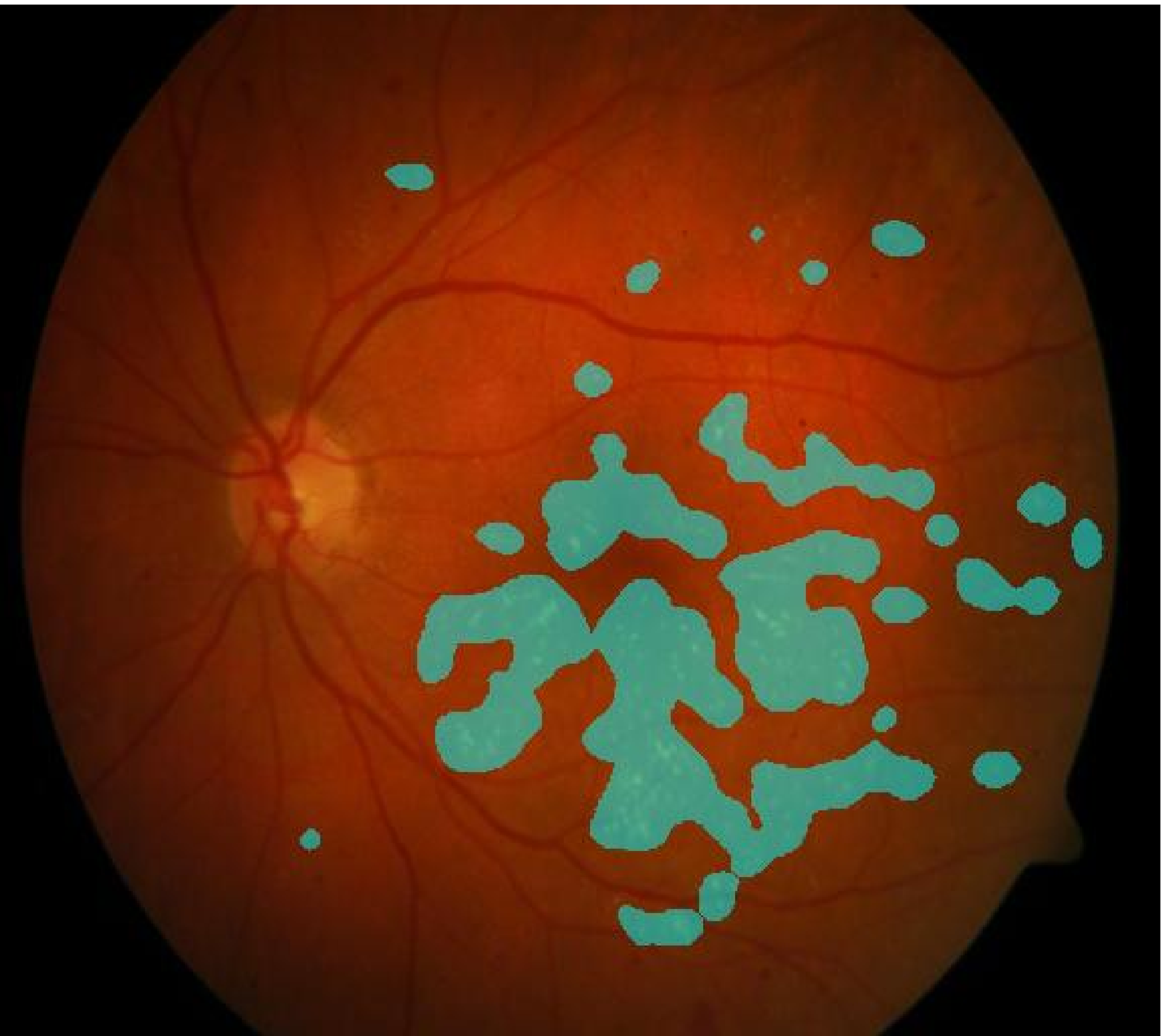}}
\subfigure[] {\label{cropping_a}\includegraphics[width=1in]{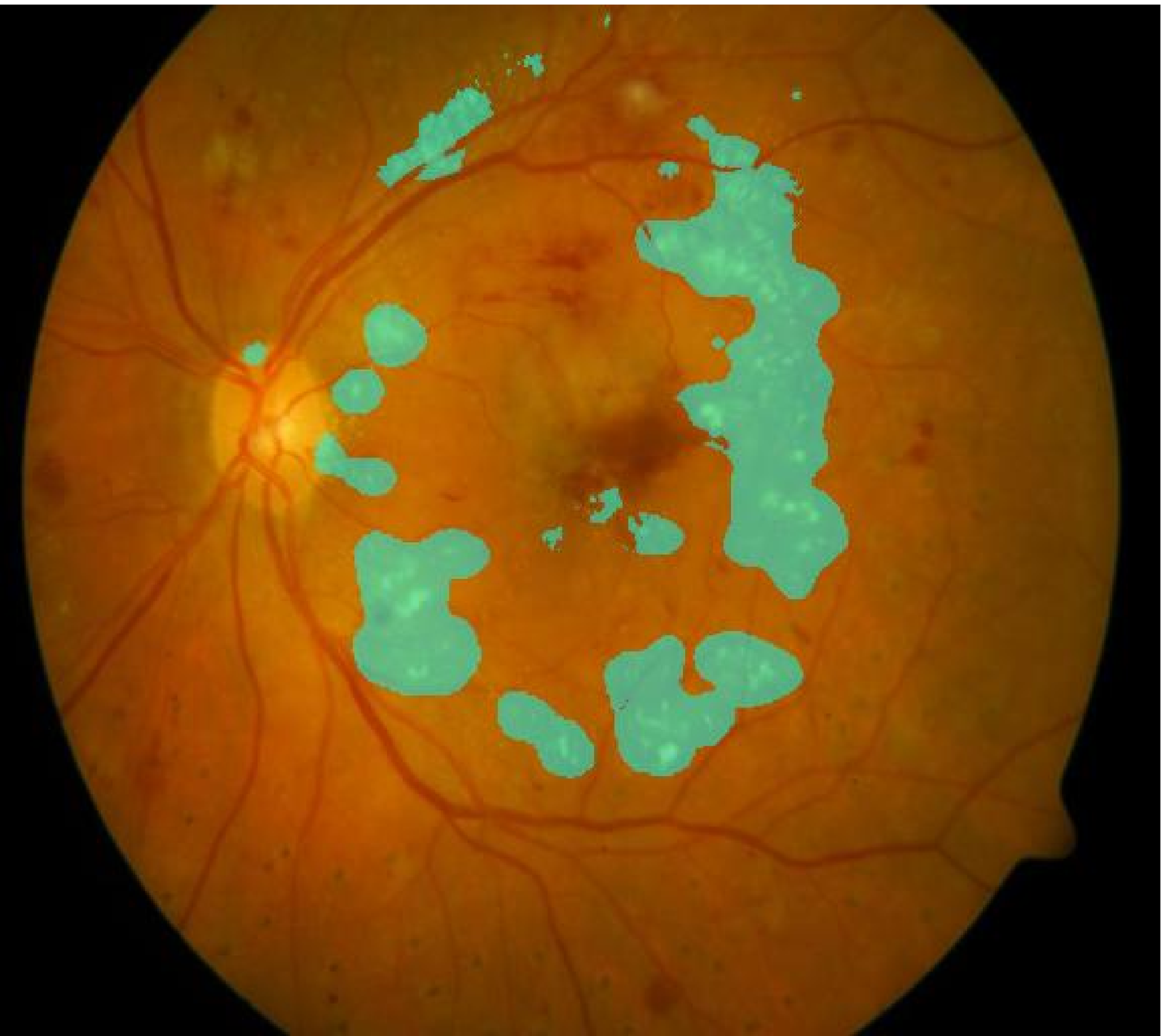}}

\subfigure[] {\label{cropping_b}\includegraphics[width=1in]{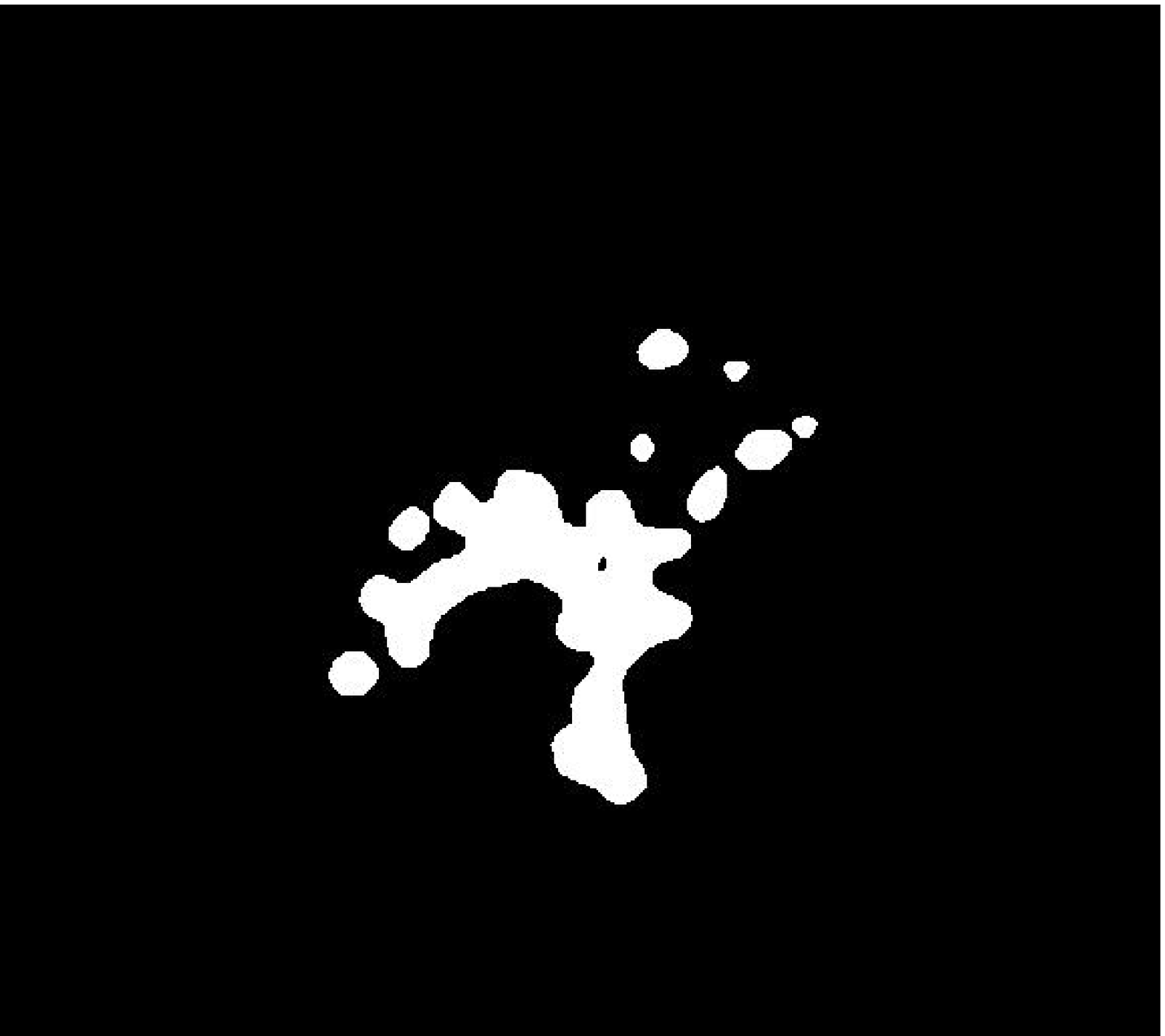}}
\subfigure[] {\label{cropping_b}\includegraphics[width=1in]{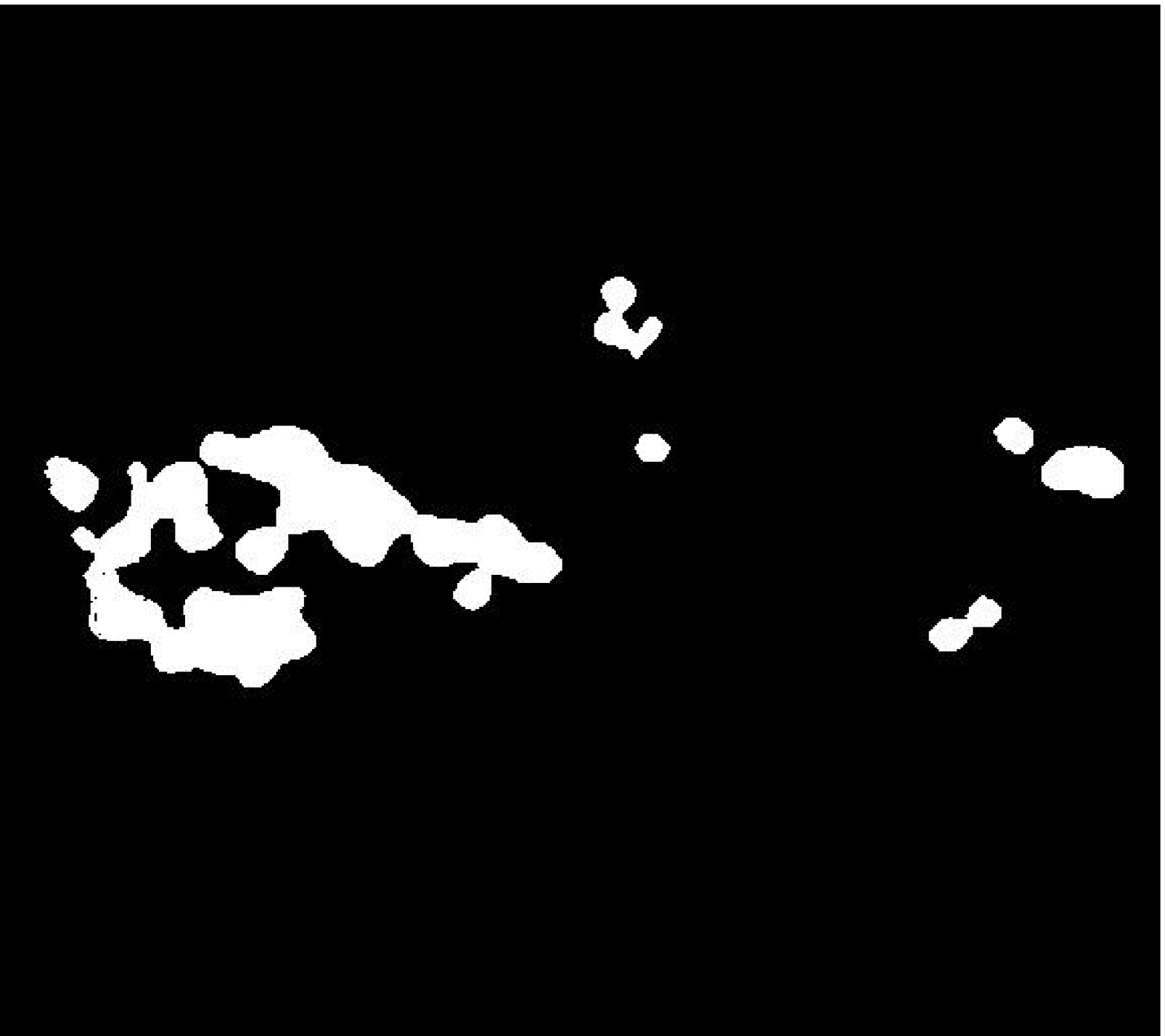}}
\subfigure[] {\label{cropping_b}\includegraphics[width=1in]{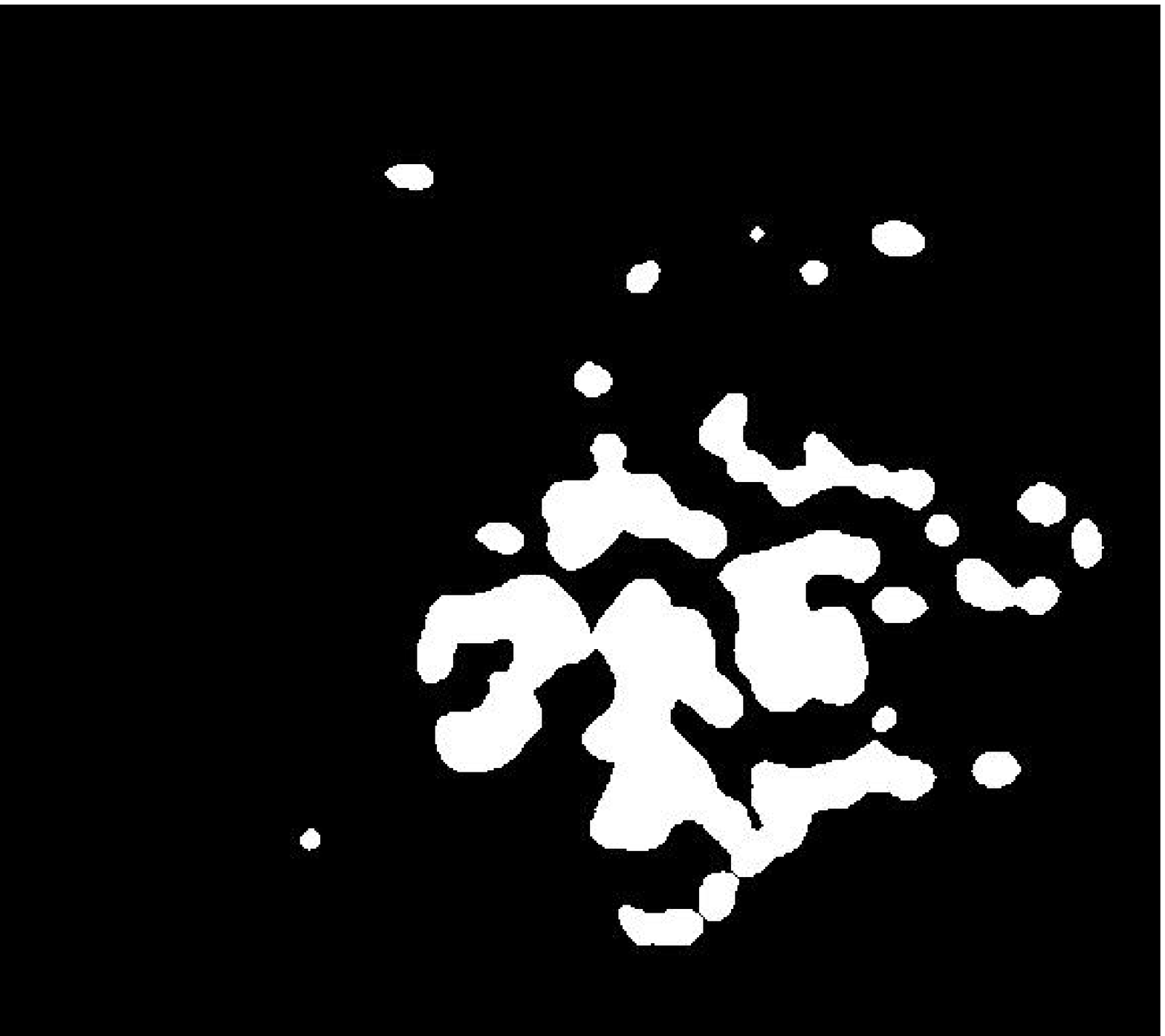}}
\subfigure[] {\label{cropping_b}\includegraphics[width=1in]{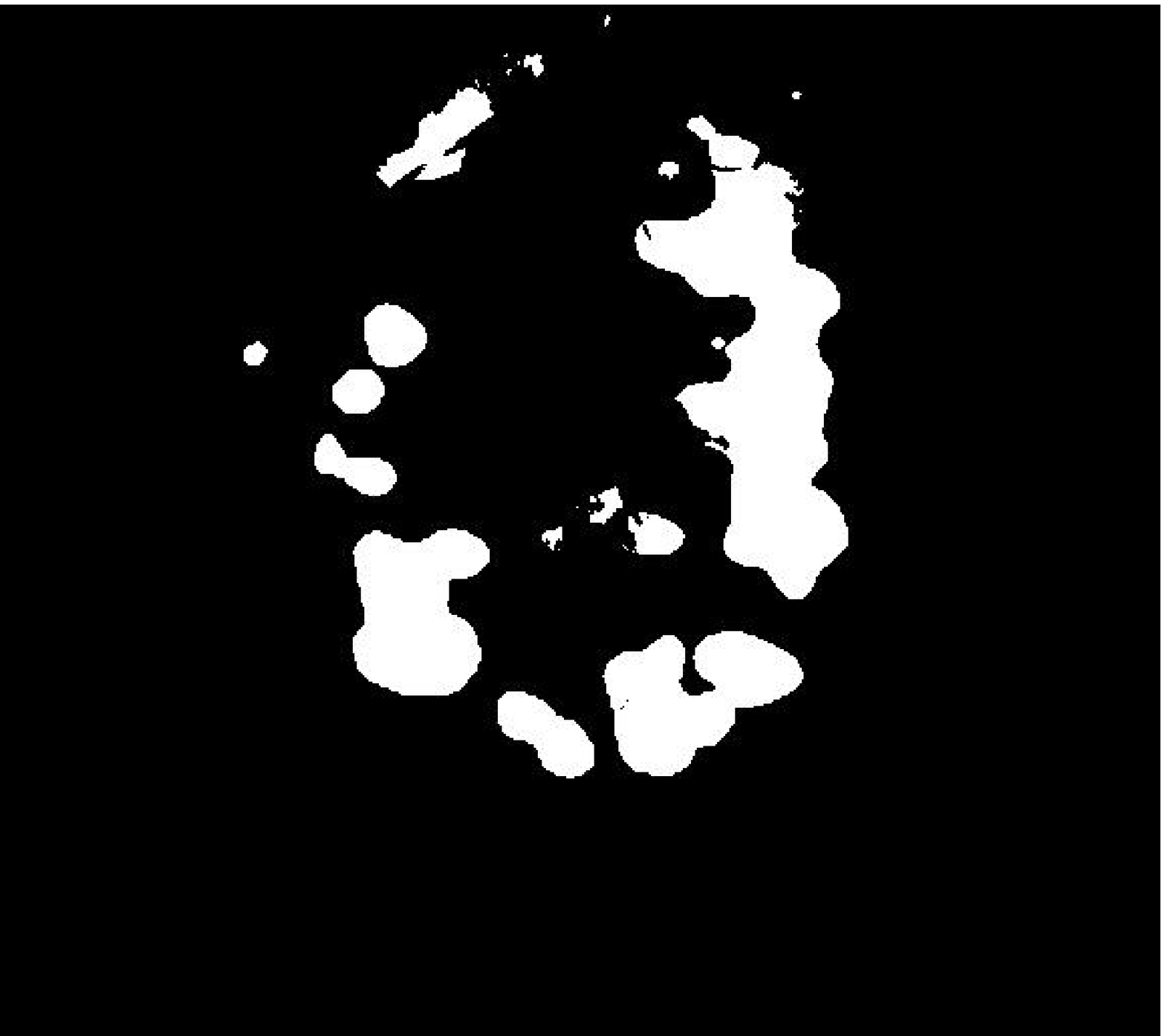}}
\caption{DiaRetDB1 database results (a) Row 1 showing the original image (b) Row 2 showing the segmented overlay (c) Row 3 showing the result of segmented image}
\label{visualD}
\end{figure*}

We employ the E-ophtha database that contains $47$ images of exudate with label ground truth and 35 healthy image. Data augmentation is performed by x and y scaling and cropping of images and $1960$ images obtained from previous $60$  images remaining $22$ images used for testing. Then training is performed with $20$ epoch and $.0001$ learning rate with $0.0005$ regularizations with a mini-batch size of $4$ images. The self-design network architecture RTC-Net has been used for this purpose having many skip connections to avoid the convolution because of more the convolution layer resulting in more the noise in the output segmented images.
\begin{table*}[!t]
\centering
\caption{Parametric details of the upsampling block and the pixel classification layer.}
\resizebox{\columnwidth}{!}{%
    \begin{tabular}{p{7em}p{2.215em}p{13.855em}p{7.5em}p{8.645em}}
    \hline
    \textbf{Block} & \textbf{Filter} & \textbf{Name (Size)} & \textbf{Activation Size} & \textbf{No. of Parameters} \bigstrut\\
    \hline
    \multicolumn{1}{c}{} & \multicolumn{1}{c}{64} & TCon-1 (4 × 4× 512) & 56 × 64× 64 & \multicolumn{1}{c}{534352} \bigstrut\\
\cline{2-5}    Up-Sampling  & \multicolumn{1}{c}{128} & TCon-2 (4 × 4× 64) & 112 × 128 × 128 & \multicolumn{1}{c}{131200} \bigstrut\\
\cline{2-5}    block & \multicolumn{1}{c}{256} & TCon-3 (4 × 4× 128) & 224 × 256 × 256 & \multicolumn{1}{c}{524544} \bigstrut\\
\cline{2-5}    \multicolumn{1}{c}{} & \multicolumn{1}{c}{512} & TCon-3 (4 × 4× 256) & 448 × 512 × 512 & \multicolumn{1}{c}{2097664} \bigstrut\\
    \hline
    Pixel   & \multicolumn{1}{c}{2} & Con-class (1 × 1 × 512) & 448 × 512 × 2 & \multicolumn{1}{c}{1026} \bigstrut\\
\cline{2-5}    classification &       & Softmax loss & 448 × 512 × 2 &   \bigstrut\\
\cline{2-5}    block &       & Pixel Classification &       &   \bigstrut\\
    \hline
    \end{tabular}%

    }
  \label{upsample}%
\end{table*}%
After training the result is evaluated on the publicly available benchmark database HEI-MED and DiaRetDB1. The exudate segmentation performance is computed against the professional expert labeled GT of DiaRetDB1 database and HEI-MED as shown in Figure.~\ref{results}. However, in the case of DiaretDB1 the GT image is provided as $4$ different grayscale images, one level according to each professional marking. Therefore, as in the case of exudate, we combined four experts' labels in one binary layer image. The segmentation results of exudates detection are given in the result section. A result is defined as TN if the segmented image does do not contains any exudate symptom both the case of GT and image. The result is defined as TP if the image, as well as GT, contain exudates. Additionally, a result is showing FP if the image does not contain exudate according to GT, but our system shows some exudates. The result showing FN if the image and GT have exudate, but our proposed algorithm does not show as given in Table.~\ref{Confusion}. The summary of our proposed semantic segmentation is shown in Figure.~\ref{visualD}.

\section{\bf \textsc{Benchmark Database}}
In the proposed method three different types of database are used as bench mark for result evaluation and comparison with previous method.
\begin{enumerate}
  \item E-ophtha \cite{decenciere2013teleophta} is a retinal color image database dedicated to scientific research on diabetic retinopathy. It was generated by the OPHDIAT© Tele-medical network for diabetic retinopathy inspection of the ANR-TECSAN-TELEOPHTA project funded by the French Research Institute (ANR). The dataset consists of two sub-directories of E-ophtha-MA’s (micro-aneurysms) and E-ophtha-EX (Exudates). The number of images used for testing and training are given in Table.~\ref{database}. The E-ophtha-EX database contains $82$ retinal images with four different dimension sizes $2048\times1360$, $1504\times1000$, $1440\times960$ and $2544\times1696$ pixels. The images are obtained $45$ degree FOV (field of view).
  \item DIARETDB1 \cite{kamarainen2007diaretdb1} is a general-purpose database for comparing diabetic retinopathy identification. The DIARETDB1 dataset contains $89$ fundus color images, of which at least $84$ contained micro-aneurysms for diabetic retinopathy and $5$ of them are normal images showed no signs of diabetic retinopathy according to all the team experts involved in the evaluation, $5$ of them were as usual and showed no signs of diabetic retinopathy. This data is compatible with the actual situation (not typical) where the images are comparable and can help to evaluate the overall performance of the diagnostic technique. This data set is called a level $1$ calibration photo.
\begin{table*}[!t]
\centering
\caption{Number of images in different data base used for different purpose}
\resizebox{\columnwidth}{!}{%
        \begin{tabular}{cccccc}
    \toprule
    \multicolumn{1}{p{2.355em}}{\textbf{Sr. No.}} & \textbf{Database} & \textbf{Total Images} & \textbf{No. Train Images} & \textbf{No. Test Images} & \textbf{Groun truth} \\
    \midrule
    1     & EOPHTHA & 82    & 60    & 22    & 82 \\
    2     & DIARETDB1 & 89    & 0     & 89    & 89 \\
    3     & HEI-MED & 169   & 0     & 169   & 169 \\
    \bottomrule
    \end{tabular}%

    }
  \label{database}%
\end{table*}%
  \item The HEI-MED \cite{akram2014automated} (Hamilton Ophthalmology Institute's Macular Edema) Database formerly known as DMED is a batch of $169$ retinal images in JPEG compressed format for testing and training by image processing algorithms to detect DME and exudates. These images were collected as part of a telemedicine network for the detection and treatment of diabetic retinopathy developed by the Hamilton Eye Institute, Machine vision team and Image Science at ORNL's with the help of collaboration of the University of Burgundy. The database was manually segmented by Dr. Edward Chaum (HEI ophthalmologist). There is no differentiation between soft and hard exudates because these differences lead to error and do not help in a clear clinical diagnostic. In addition to GT and images, the database provides anonymous clinical data about the patient, manually detected optic nerve locations, blood vessels segmented in the machine, and Matlab classes for easy access to all data and elements, without having to deal with internal file formats.
\end{enumerate}

\section{\bf \textsc{Experimental analysis}}
In this section, the results of supervised neural architecture are tested on the $3$ publicly available retinal database E-ophtha, HEI-MED and DiaRetDB1 to test the performance of the proposed method.

\subsection{\bf \textsc{Evaluation measures}}
\label{sec:6.3}
Five measures are utilized in this proposed work for quantitative analysis including area overlap, specificity, sensitivity, accuracy and Dice similarity-coefficient. The quantitative calculation are computed as a result of confusion matrix can be seen in Table~\ref{Confusion}. The quantities analysis to understand the confusion matrix are given below:

\begin{equation}\label{Precision}
   Precision=\frac{TP}{(TP+FP)}
 \end{equation} \color{black}
Sensitivity (SN) is ratio of true positive and is known as $recall$. It computes the part of correctly segmented TP as true, given as
  \begin{equation}\label{Sensitivity}
   SN=\frac{TP}{FN+TP}.
 \end{equation}

The ratio of true negative known as specificity that means the value of negatives that are segmented as negatives. It is measured as:

 \begin{equation}\label{Specificity}
   SP=\frac{TN}{TN+FP}.
 \end{equation}\color{black}
The ratio of truly segmented observation is called accuracy.
 \begin{equation}\label{Accuracy}
  Accuracy=\frac{TN+TP}{TN+TP+FP+FN}
 \end{equation}

\begin{figure*}[!t]
\centering
\subfigure[] {\label{cropping_a}\includegraphics[width=1in]{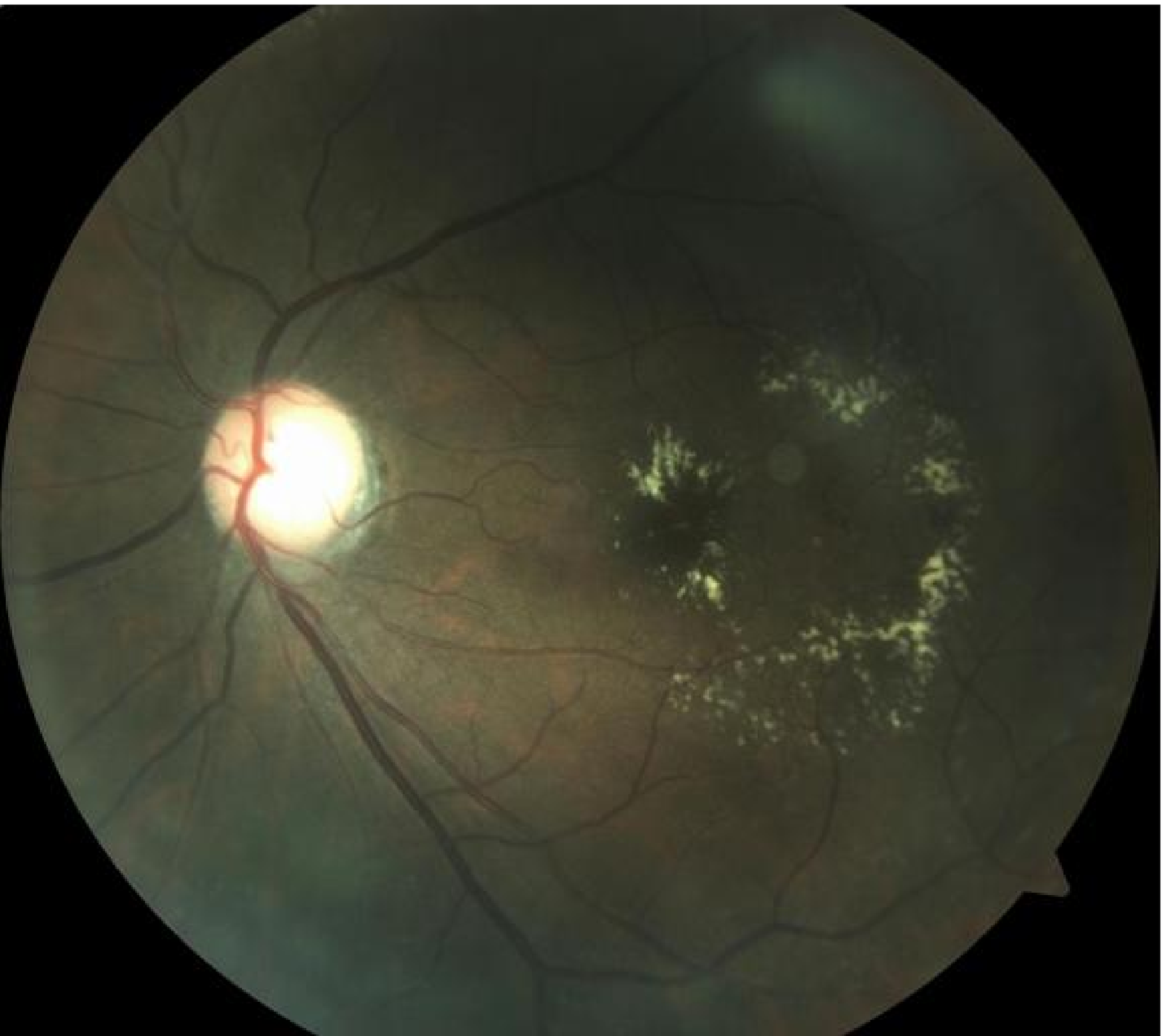}}
\subfigure[] {\label{cropping_a}\includegraphics[width=1in]{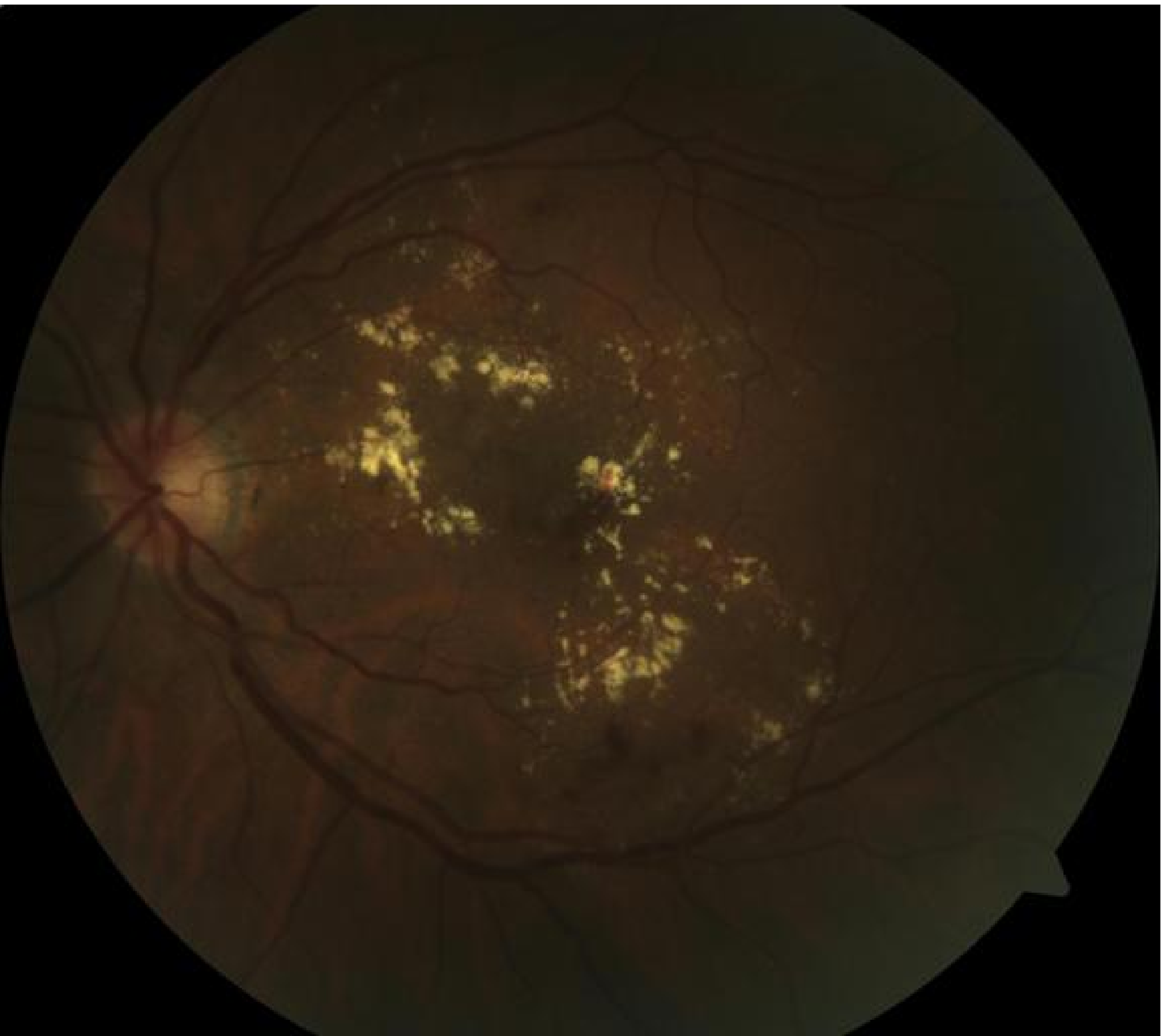}}
\subfigure[] {\label{cropping_a}\includegraphics[width=1in]{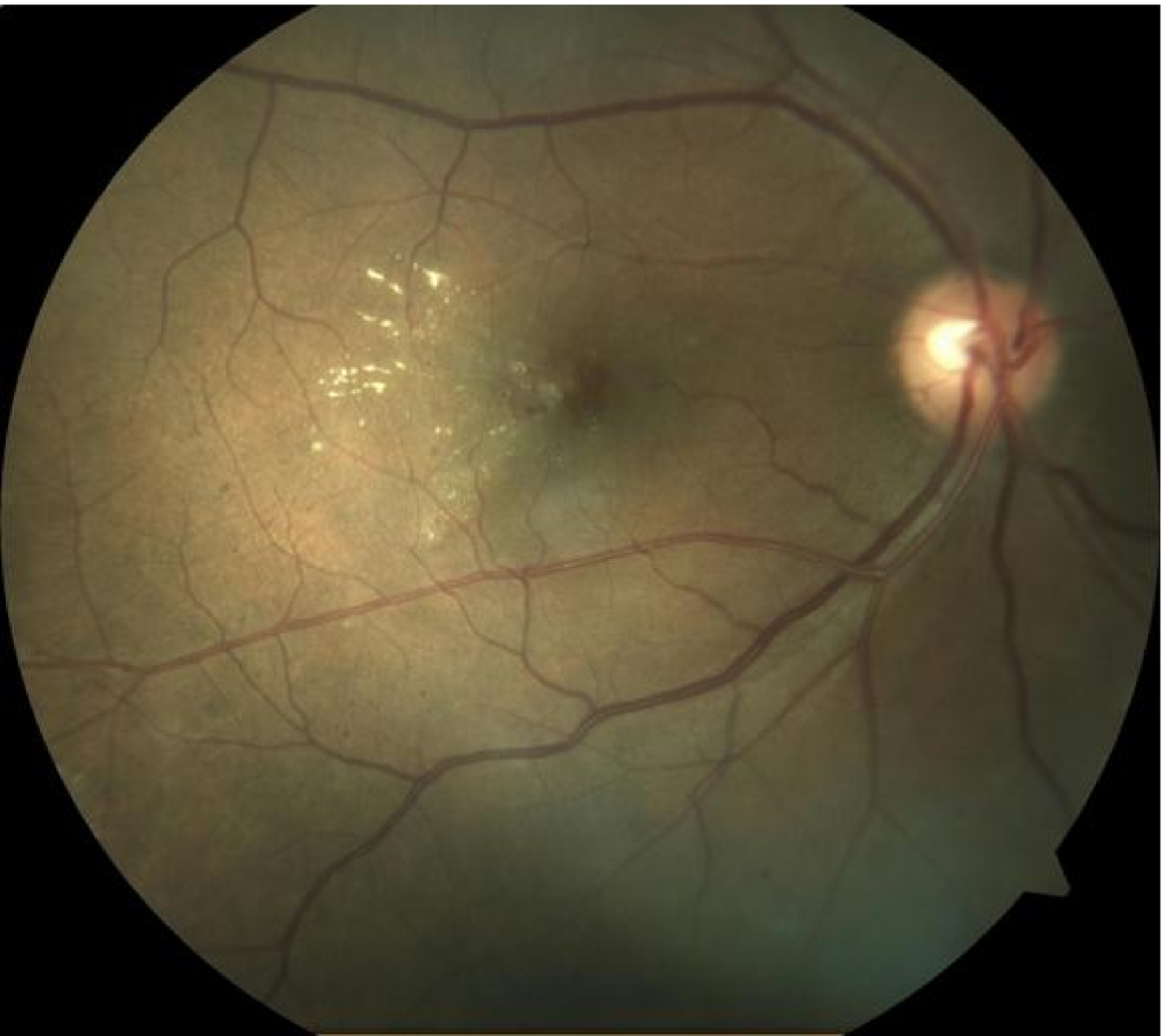}}
\subfigure[] {\label{cropping_a}\includegraphics[width=1in]{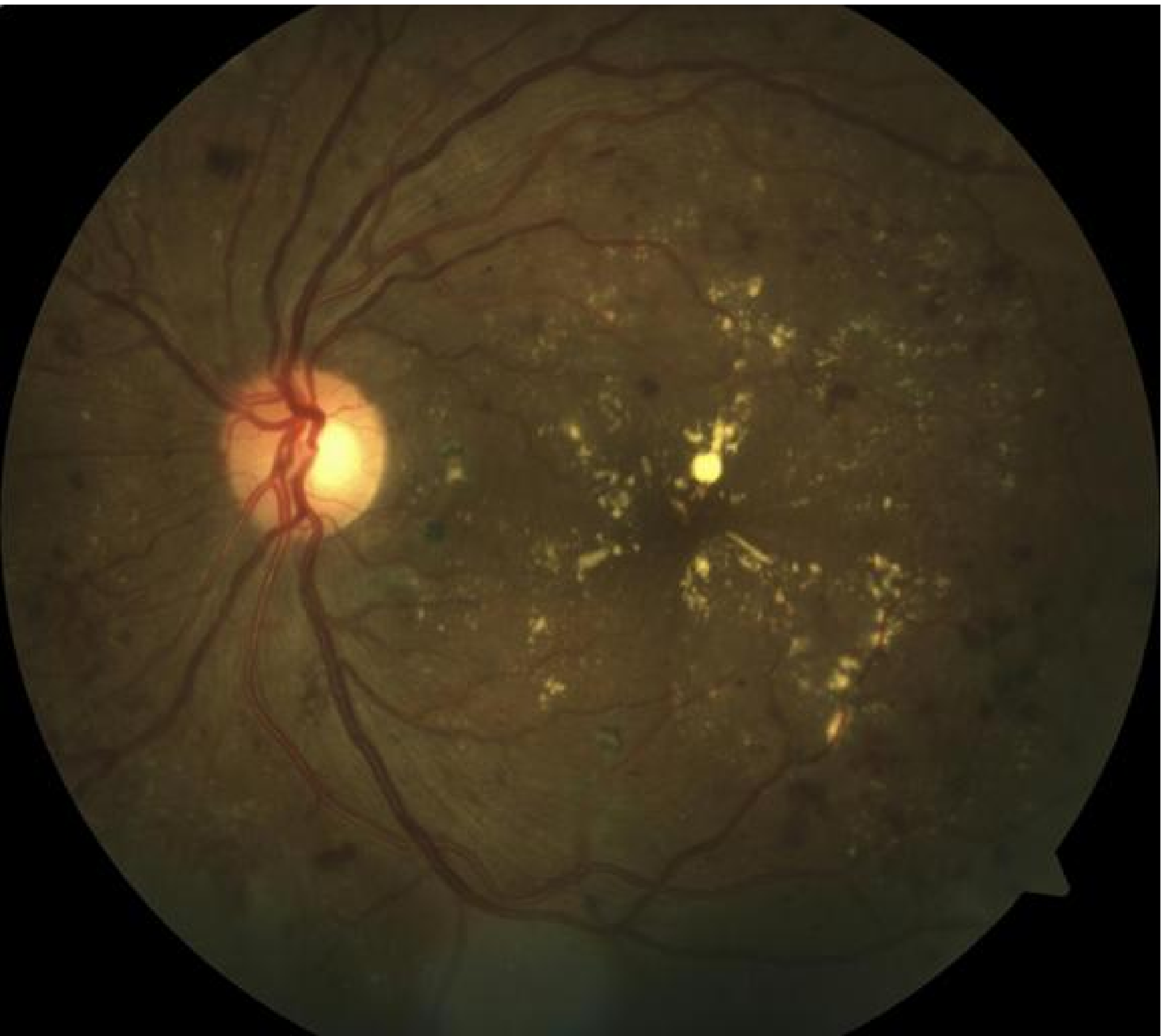}}

\subfigure[] {\label{cropping_a}\includegraphics[width=1in]{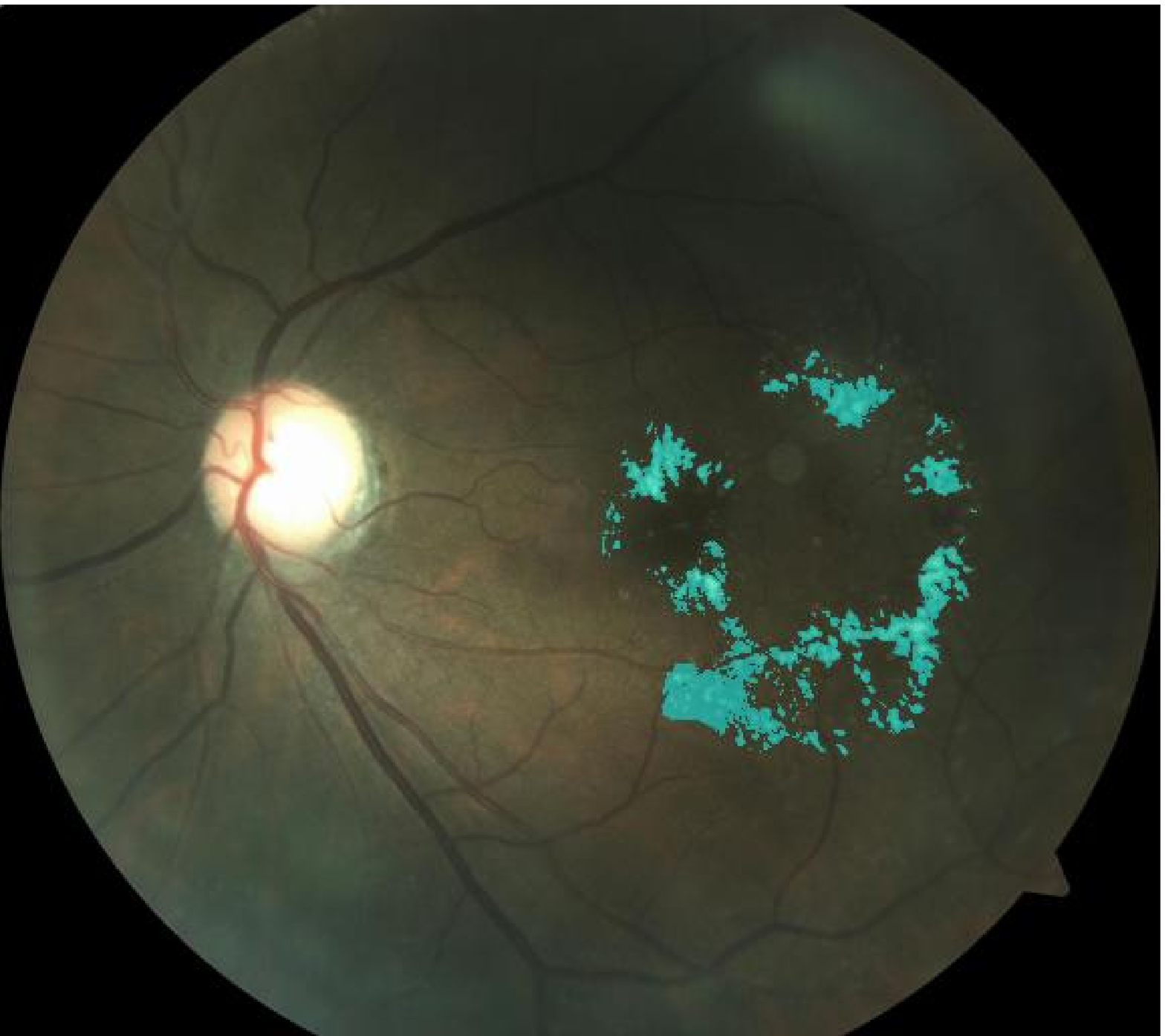}}
\subfigure[] {\label{cropping_a}\includegraphics[width=1in]{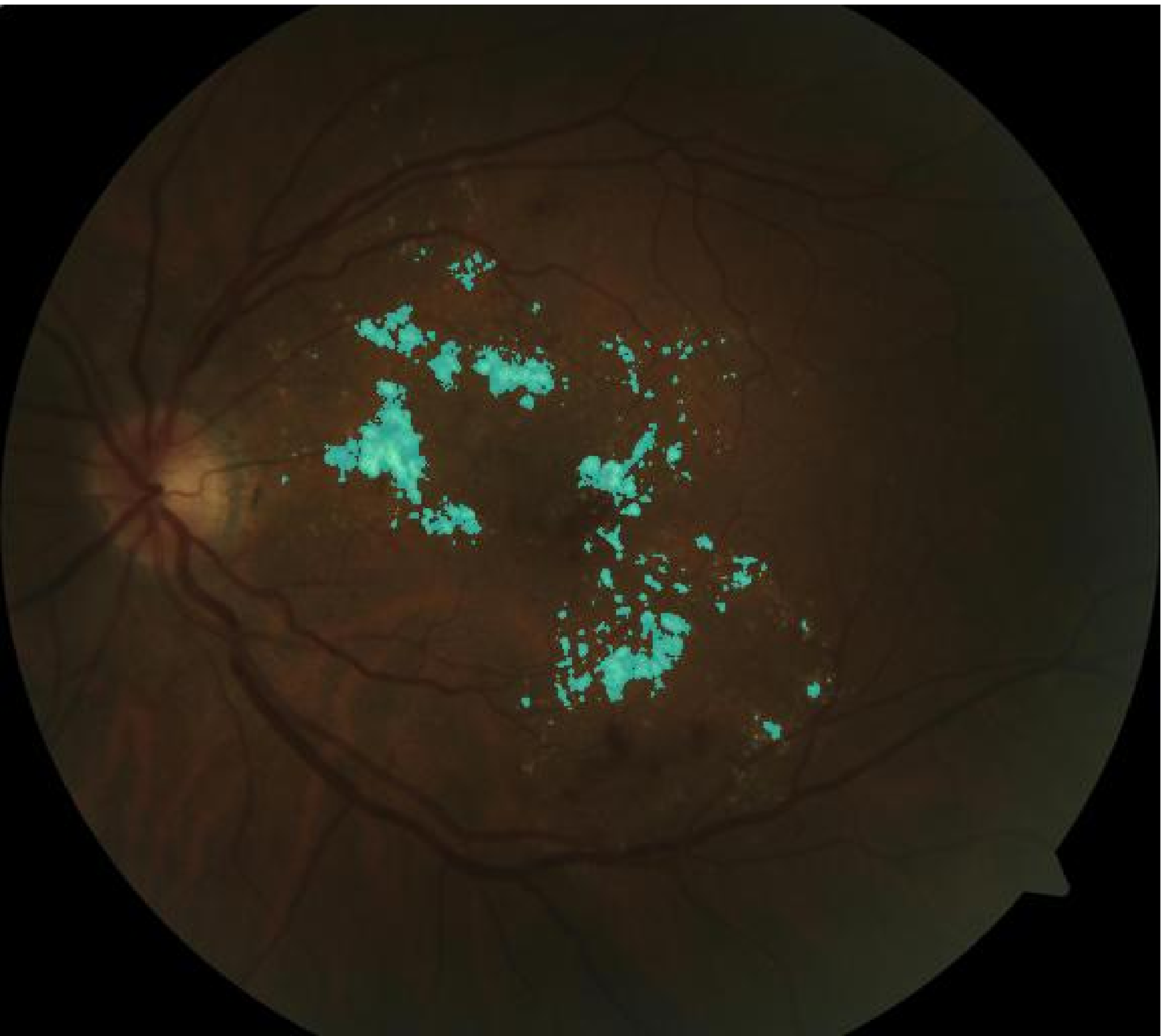}}
\subfigure[] {\label{cropping_a}\includegraphics[width=1in]{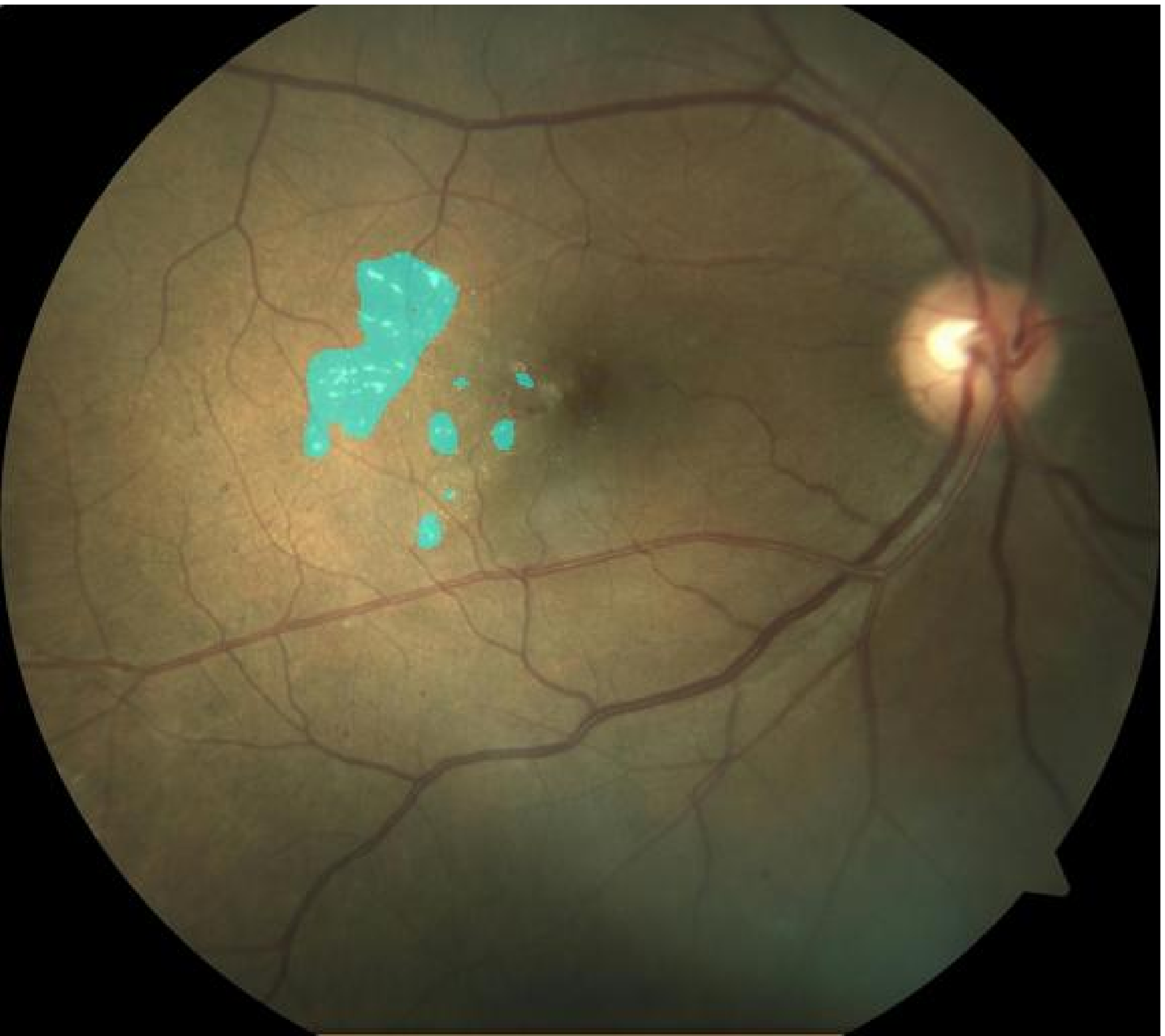}}
\subfigure[] {\label{cropping_a}\includegraphics[width=1in]{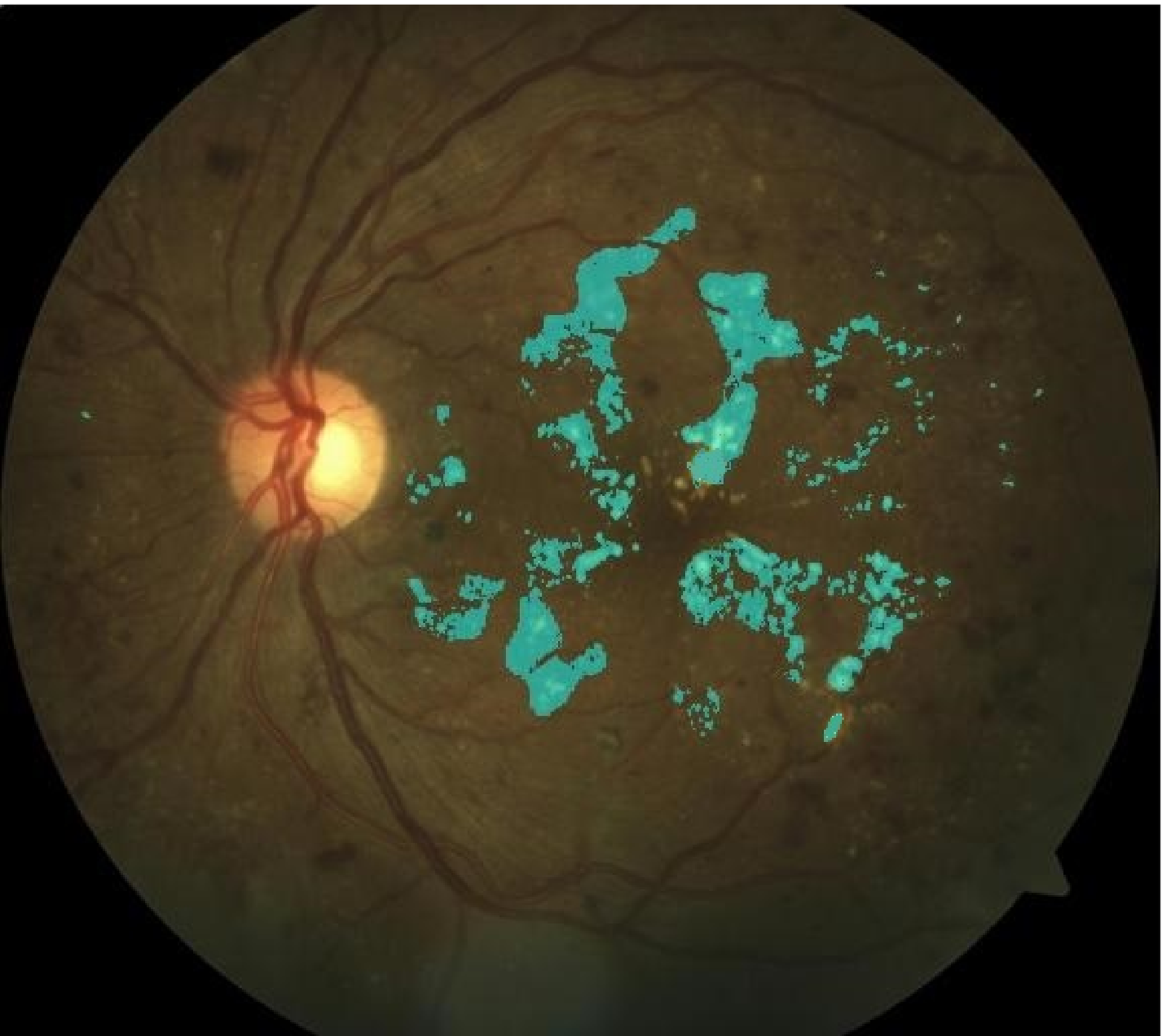}}

\subfigure[] {\label{cropping_b}\includegraphics[width=1in]{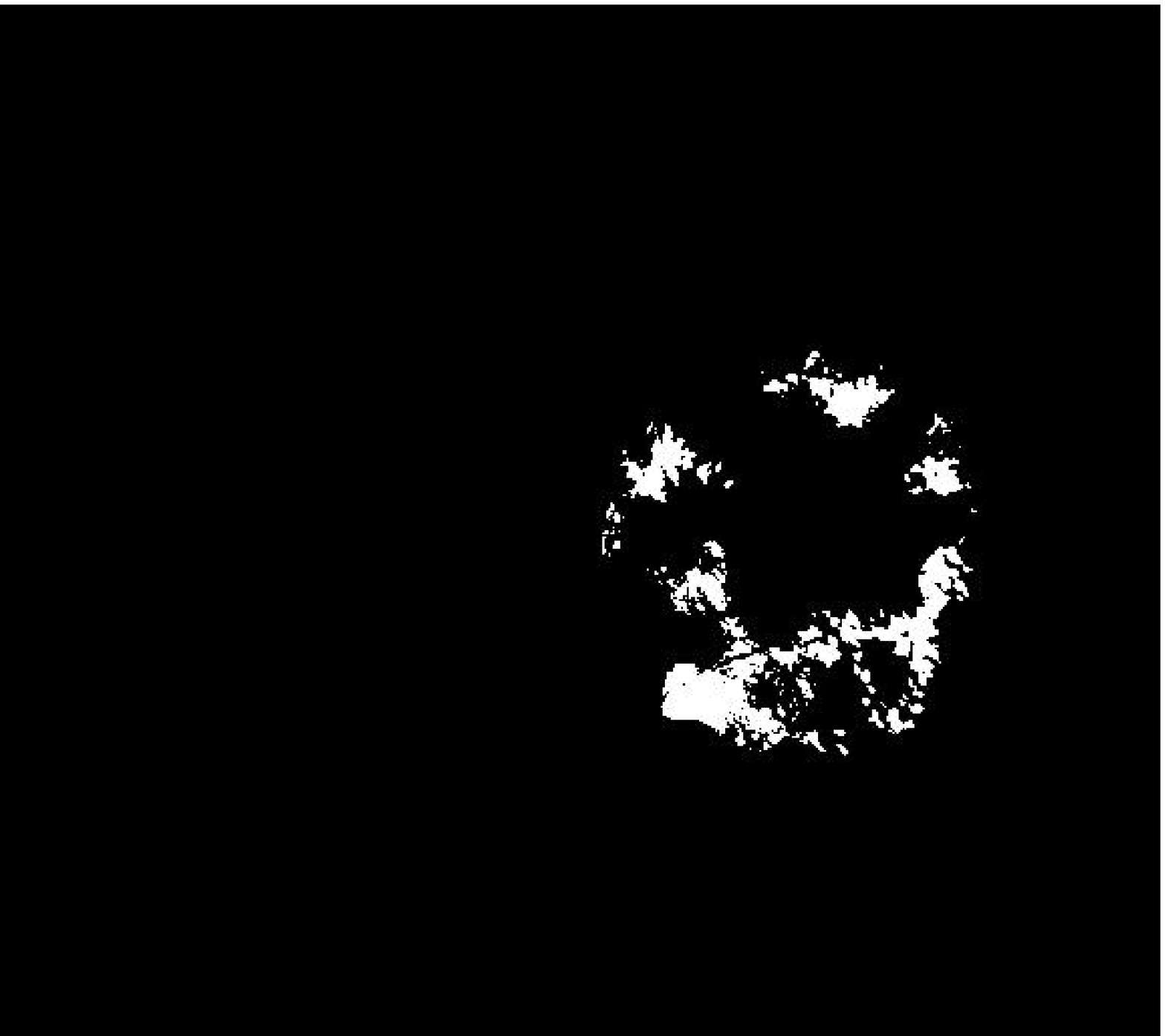}}
\subfigure[] {\label{cropping_b}\includegraphics[width=1in]{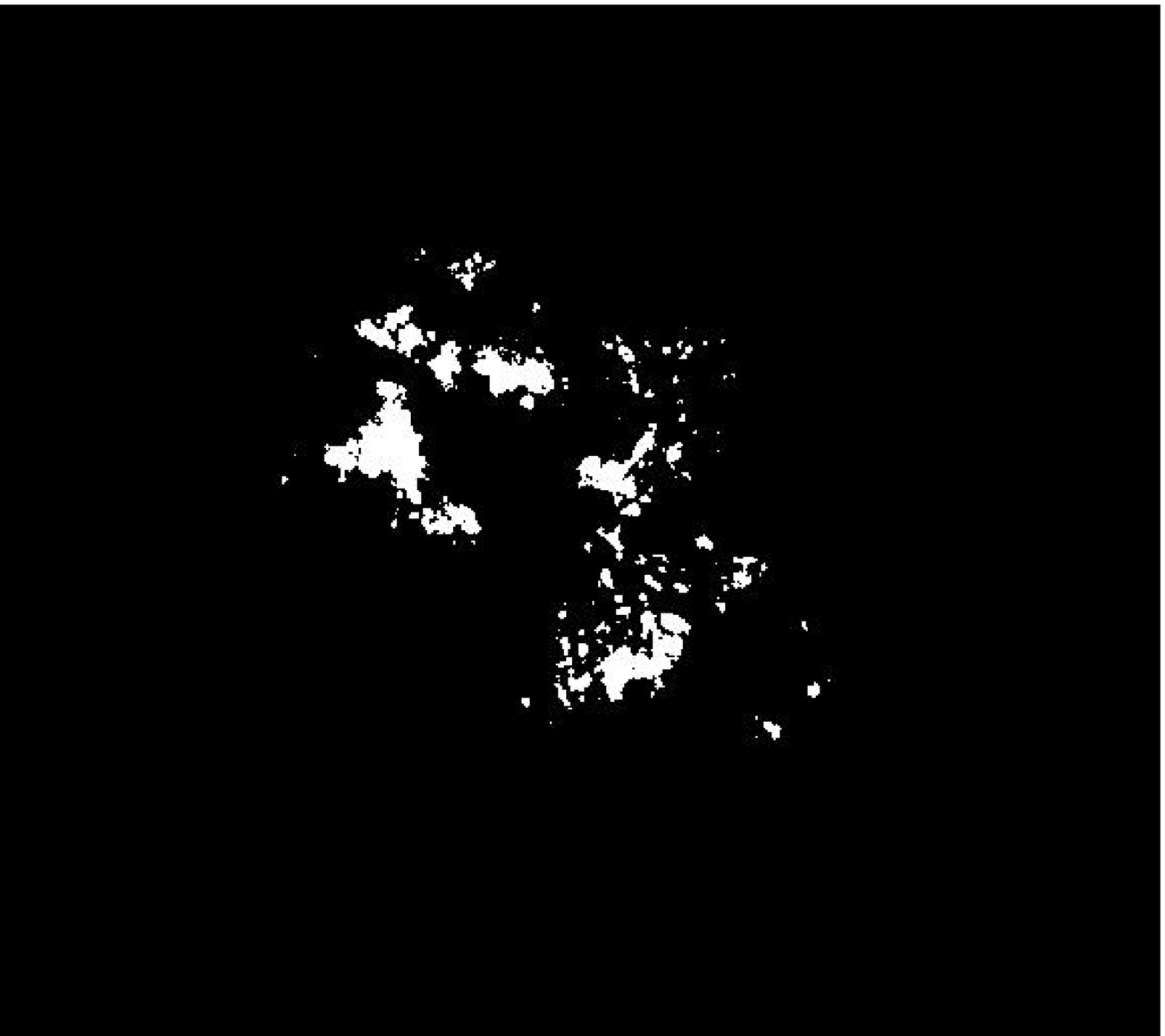}}
\subfigure[] {\label{cropping_b}\includegraphics[width=1in]{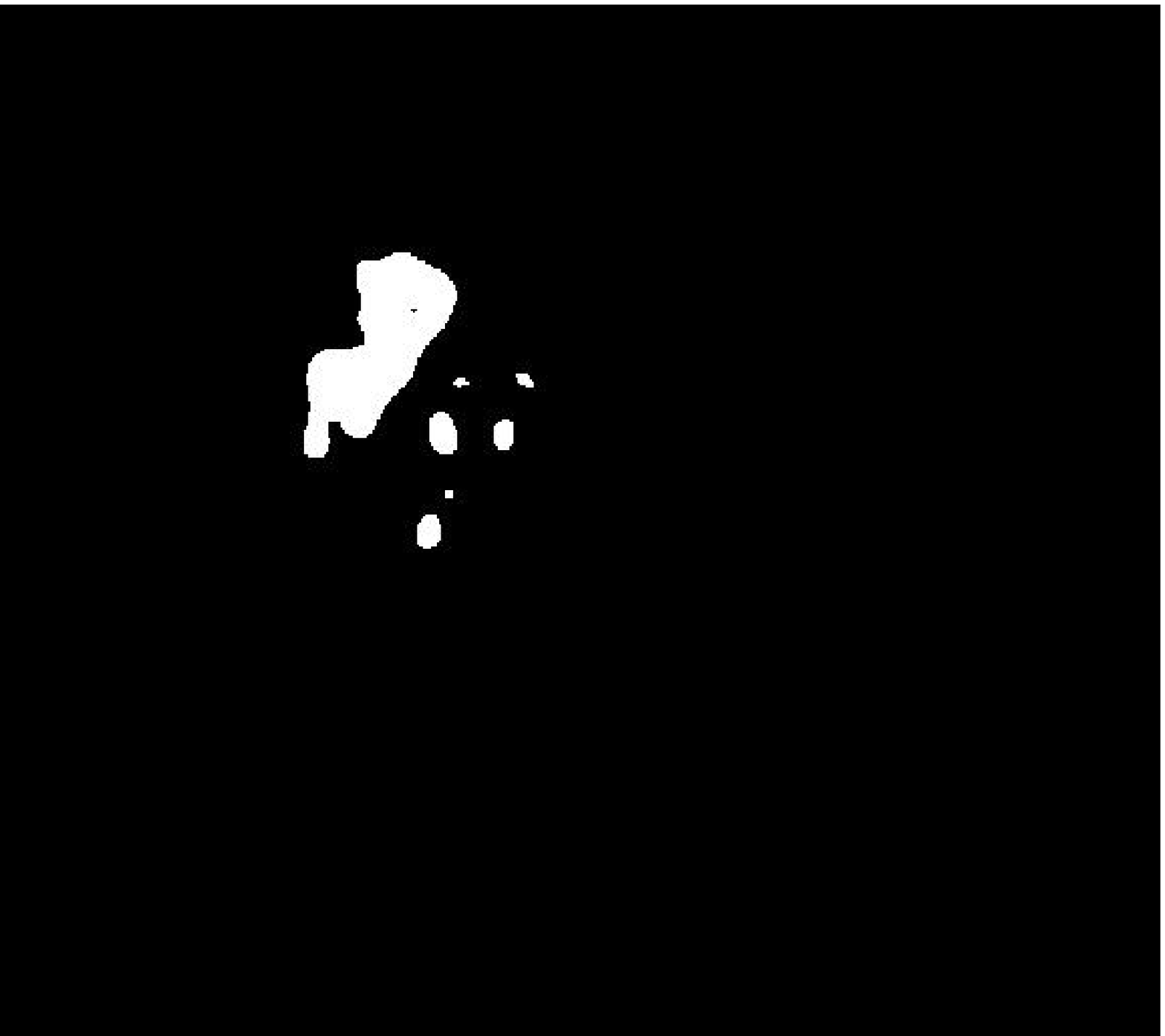}}
\subfigure[] {\label{cropping_b}\includegraphics[width=1in]{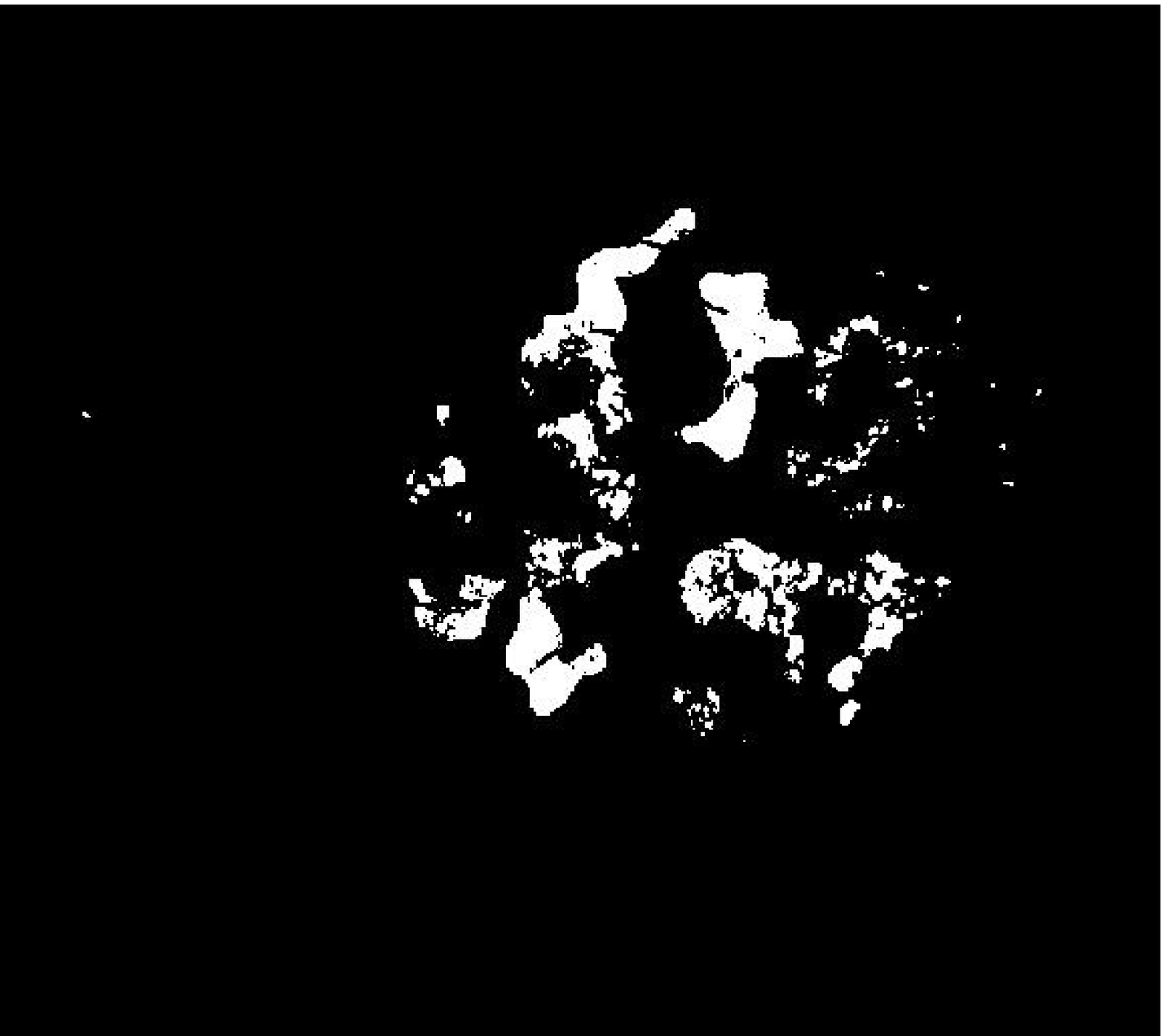}}
\caption{HEI-MED database results (a) Row 1 showing the original image (b) Row 2 showing the segmented overlay (c) Row 3 showing the result of segmented image}
\label{visualH}
\end{figure*}

\begin{table}
\caption{Confusion Matrix.}
   \begin{tabular}{p{7.855em}|p{5.645em}|p{5.57em}}
    \multicolumn{1}{r|}{} & Exudate & Non-Exudate  \bigstrut[b]\\
    \hline
    Classify correctly & True Positive (TP) & False Positive (FP) \bigstrut[t]\\
   Classify wrongly  & False Negative (FN) & True Negative (TN) \\
    \end{tabular}% 
  \label{Confusion}
\end{table}
\begin{itemize}
  \item TP: correctly segmented exudates pixels.
  \item FP: incorrectly segmented non-exudates pixels.
  \item TN: correctly segmented Non-exudates pixels.
  \item FN: incorrectly exudates  pixels as non-exudates pixels.
\end{itemize}
The overall comparison of proposed method with other State-of-art is given in Table \ref{table6} and result of HEI-MED as cross validation is given in Figure \ref{visualH}.
\begin{table*}[!t]
\resizebox{\columnwidth}{!}{%
\caption{Comparison with the state-of-the-art.}

     \begin{tabular}{cp{12.715em}p{4.645em}p{4.93em}p{5em}p{5.5em}p{4.355em}}
    \toprule
    \multicolumn{1}{p{1.855em}}{\textbf{Sr. No.}} & \multicolumn{1}{l}{\textbf{Performance measures}} & \multicolumn{1}{c}{\textbf{Precision}} & \multicolumn{1}{c}{\textbf{Accuracy}} & \multicolumn{1}{c}{\textbf{Sensitivity}} & \multicolumn{1}{c}{\textbf{Specificity}} & \multicolumn{1}{c}{\textbf{AUC}} \\
    \midrule
          & \multicolumn{1}{l}{\textbf{Method }} & \multicolumn{1}{r}{} & \multicolumn{1}{r}{} & \multicolumn{1}{r}{} & \multicolumn{1}{r}{} & \multicolumn{1}{r}{} \\
\cmidrule{1-2}          & \multicolumn{1}{l}{DIARETDB1} & \multicolumn{1}{r}{} & \multicolumn{1}{r}{} & \multicolumn{1}{r}{} & \multicolumn{1}{r}{} & \multicolumn{1}{r}{} \\
    \midrule
    1     & \cite{imani2016novel} & \multicolumn{1}{c}{\textcolor[rgb]{ .137,  .122,  .125}{0.82}} & \textcolor[rgb]{ .137,  .122,  .125}{~~~~~---~} & \textcolor[rgb]{ .137,  .122,  .125}{~~~~~0.89 } & \multicolumn{1}{c}{\textcolor[rgb]{ .137,  .122,  .125}{0.99}} & \multicolumn{1}{c}{\textcolor[rgb]{ .137,  .122,  .125}{0.96}} \\
    2     & \cite{akram2012automated} & \multicolumn{1}{c}{\textcolor[rgb]{ .137,  .122,  .125}{0.97}} & \textcolor[rgb]{ .137,  .122,  .125}{~~~~~---~} & \multicolumn{1}{c}{\textcolor[rgb]{ .137,  .122,  .125}{0.93}} & \multicolumn{1}{c}{\textcolor[rgb]{ .137,  .122,  .125}{0.99}} & \textcolor[rgb]{ .137,  .122,  .125}{~~~~~---~} \\
    3     & \cite{fraz2017multiscale} & \multicolumn{1}{c}{0.87} & \multicolumn{1}{c}{0.87} & \multicolumn{1}{c}{0.92} & \multicolumn{1}{c}{0.81} & \multicolumn{1}{c}{0.93} \\
    4     & \cite{almotiri2018multi} & \multicolumn{1}{c}{0.42} & \multicolumn{1}{c}{\textcolor[rgb]{ .137,  .122,  .125}{0.83}} & \multicolumn{1}{c}{0.75} & \multicolumn{1}{c}{0.85} & \textcolor[rgb]{ .137,  .122,  .125}{~~~~~---~} \\
    5     & \cite{liu2017location} & {~~~~~---~}     & \multicolumn{1}{c}{\textcolor[rgb]{ .137,  .122,  .125}{0.79}} & \multicolumn{1}{c}{0.83} & \multicolumn{1}{c}{0.75} & \textcolor[rgb]{ .137,  .122,  .125}{~~~~~---~} \\
    6     & \cite{zheng2018detection} & \multicolumn{1}{c}{0.91} & \multicolumn{1}{c}{0.99} & \multicolumn{1}{c}{0.93} & \multicolumn{1}{c}{0.99} & \textcolor[rgb]{ .137,  .122,  .125}{~~~~~---~} \\
    7     & \textbf{Proposed} & \textbf{~~~~\textbf{0.92}~} & \textbf{~~~~~\textbf{0.98}~} & \textbf{~~~~~\textbf{0.95}~} & \textbf{~~~~~~\textbf{0.99}~} & \textbf{~~~~0.94~} \\
    \midrule
          & \multicolumn{1}{l}{E-OPHTHA} & \multicolumn{1}{r}{} & \multicolumn{1}{r}{} & \multicolumn{1}{r}{} & \multicolumn{1}{r}{} & \multicolumn{1}{r}{} \\
    \midrule
    1     & \cite{imani2016novel} & \multicolumn{1}{c}{\textcolor[rgb]{ .137,  .122,  .125}{0.77}} & \textcolor[rgb]{ .137,  .122,  .125}{~~~~~---~} & \multicolumn{1}{c}{\textcolor[rgb]{ .137,  .122,  .125}{0.8}} & \multicolumn{1}{c}{\textcolor[rgb]{ .137,  .122,  .125}{0.99}} & \multicolumn{1}{c}{\textcolor[rgb]{ .137,  .122,  .125}{0.93}} \\
    2     & \textcolor[rgb]{ .137,  .122,  .125}{\cite{liu2017location}} & \multicolumn{1}{c}{\textcolor[rgb]{ .137,  .122,  .125}{0.75}} & \textcolor[rgb]{ .137,  .122,  .125}{~~~~~---~} & \multicolumn{1}{c}{\textcolor[rgb]{ .137,  .122,  .125}{0.76}} & \textcolor[rgb]{ .137,  .122,  .125}{~~~~~---~} & \textcolor[rgb]{ .137,  .122,  .125}{~~~~~---~} \\
    3     & \cite{zhang2014exudate} & \multicolumn{1}{c}{0.79} & \textcolor[rgb]{ .137,  .122,  .125}{~~~~~---~} & \multicolumn{1}{c}{0.74} & \textcolor[rgb]{ .137,  .122,  .125}{~~~~~---~} & \multicolumn{1}{c}{0.95} \\
    4     & \textcolor[rgb]{ .137,  .122,  .125}{  \cite{harangi2014detection}} & \multicolumn{1}{c}{\textcolor[rgb]{ .137,  .122,  .125}{0.65}} & \textcolor[rgb]{ .137,  .122,  .125}{~~~~~---~} & \multicolumn{1}{c}{\textcolor[rgb]{ .137,  .122,  .125}{0.66}} & \textcolor[rgb]{ .137,  .122,  .125}{~~~~~---~} & \textcolor[rgb]{ .137,  .122,  .125}{~~~~~---~} \\
    5     & \cite{fraz2017multiscale} & \multicolumn{1}{c}{0.91} & \multicolumn{1}{c}{0.89} & \multicolumn{1}{c}{0.81} & \multicolumn{1}{c}{0.94} & \multicolumn{1}{c}{0.94} \\
    6     & \cite{zheng2018detection} & \multicolumn{1}{c}{0.94} & \multicolumn{1}{c}{0.99} & \multicolumn{1}{c}{0.9} & ~~~~~0.99  & \textcolor[rgb]{ .137,  .122,  .125}{~~~~~---~} \\
    \multicolumn{1}{c}{7} & \textbf{Proposed} & \textcolor[rgb]{ .137,  .122,  .125}{\textbf{~~~~~0.96~}} & \textcolor[rgb]{ .137,  .122,  .125}{\textbf{~~~~~0.99~}} & \textcolor[rgb]{ .137,  .122,  .125}{\textbf{~~~~~0.92~}} & \textcolor[rgb]{ .137,  .122,  .125}{\textbf{~~~~~0.99~}} & \textcolor[rgb]{ .137,  .122,  .125}{\textbf{~~~~0.97~}} \\
    \midrule
          & \multicolumn{1}{l}{HEI-MED} & \multicolumn{1}{r}{} & \multicolumn{1}{r}{} & \multicolumn{1}{r}{} & \multicolumn{1}{r}{} & \multicolumn{1}{r}{} \\
    \midrule
    1     & \cite{imani2016novel} & \multicolumn{1}{c}{0.63} & \textbf{~~~~~---~} & \multicolumn{1}{c}{0.81} & \multicolumn{1}{c}{0.99} & \multicolumn{1}{c}{0.94} \\
    2     & \cite{harangi2014detection} & \textcolor[rgb]{ .137,  .122,  .125}{~~~~~---~} & \multicolumn{1}{c}{0.86} & \multicolumn{1}{c}{0.87} & \multicolumn{1}{c}{0.86} & \textcolor[rgb]{ .137,  .122,  .125}{~~~~~---~} \\
    3     & \cite{zhang2014exudate} & \textcolor[rgb]{ .137,  .122,  .125}{~~~~~---~} & \textcolor[rgb]{ .137,  .122,  .125}{~~~~~---~} & \textcolor[rgb]{ .137,  .122,  .125}{~~~~~---~} & \textcolor[rgb]{ .137,  .122,  .125}{~~~~~---~} & \multicolumn{1}{c}{0.94} \\
    4     & \cite{fraz2017multiscale} & \multicolumn{1}{c}{0.93} & \multicolumn{1}{c}{0.95} & \multicolumn{1}{c}{0.94} & \multicolumn{1}{c}{0.96} & \multicolumn{1}{c}{0.98} \\
    5     & \cite{akram2014automated} & \multicolumn{1}{c}{} & \multicolumn{1}{c}{0.92} & \multicolumn{1}{c}{0.96} & \multicolumn{1}{c}{0.94} & \textcolor[rgb]{ .137,  .122,  .125}{~~~~~---~} \\
    6     & \cite{zheng2018detection} & \multicolumn{1}{c}{0.91} & \multicolumn{1}{c}{0.99} & \multicolumn{1}{c}{0.9} & \multicolumn{1}{c}{0.99} & \textcolor[rgb]{ .137,  .122,  .125}{~~~~~---~} \\
    7     & \textbf{Proposed} & \textcolor[rgb]{ .137,  .122,  .125}{\textbf{~~~~0.92~}} & \textcolor[rgb]{ .137,  .122,  .125}{\textbf{~~~~~0.98~}} & \textcolor[rgb]{ .137,  .122,  .125}{\textbf{~~~~~0.97~}} & \textcolor[rgb]{ .137,  .122,  .125}{\textbf{~~~~~0.99~}} & \textcolor[rgb]{ .137,  .122,  .125}{\textbf{~~~~0.96~}} \\
    \bottomrule
    \end{tabular}%

 }
 \label{table6}
\end{table*}

\begin{figure*}[!t]
\centering
\subfigure[] {\label{cropping_a}\includegraphics[width=1in]{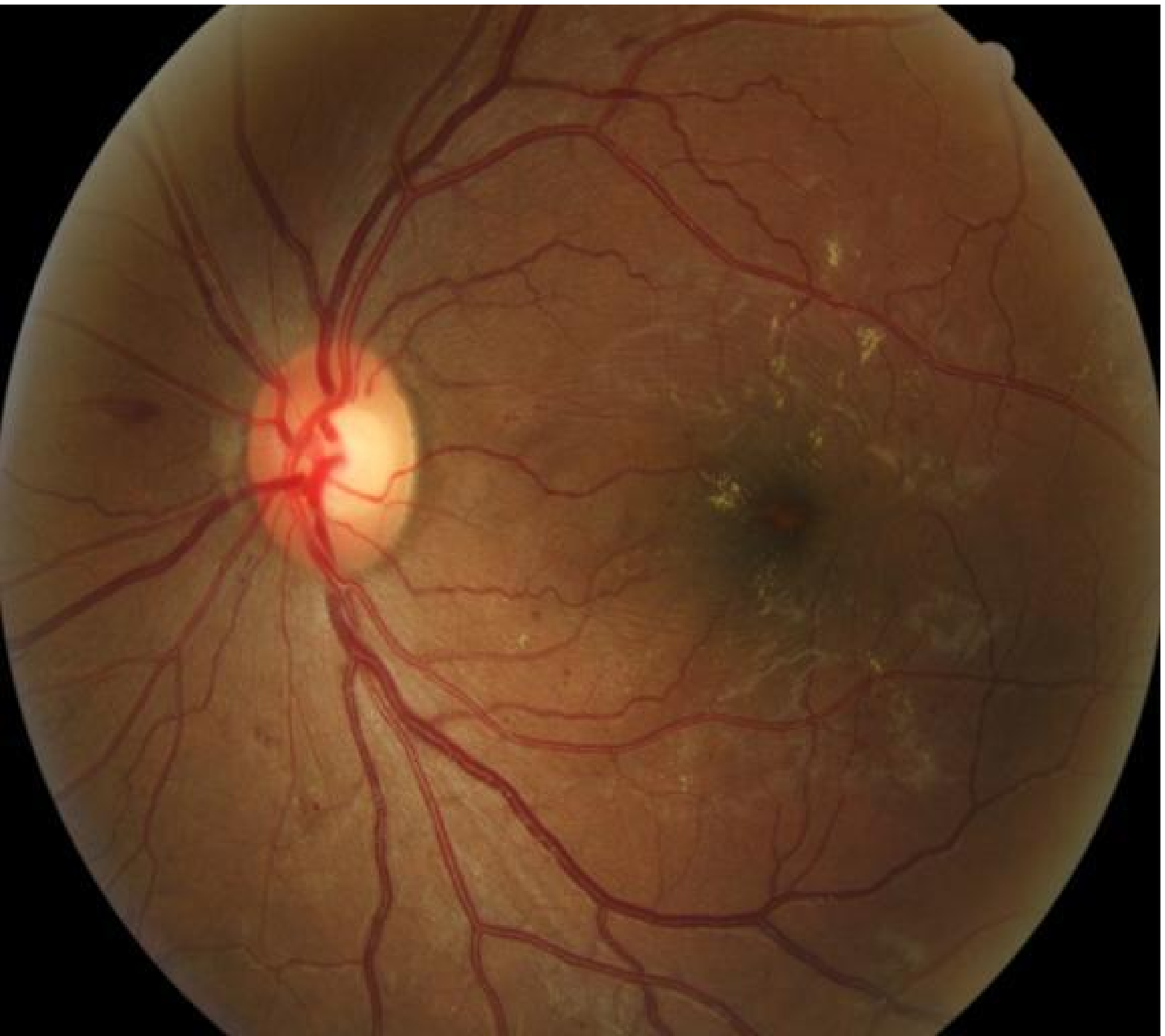}}
\subfigure[] {\label{cropping_a}\includegraphics[width=1in]{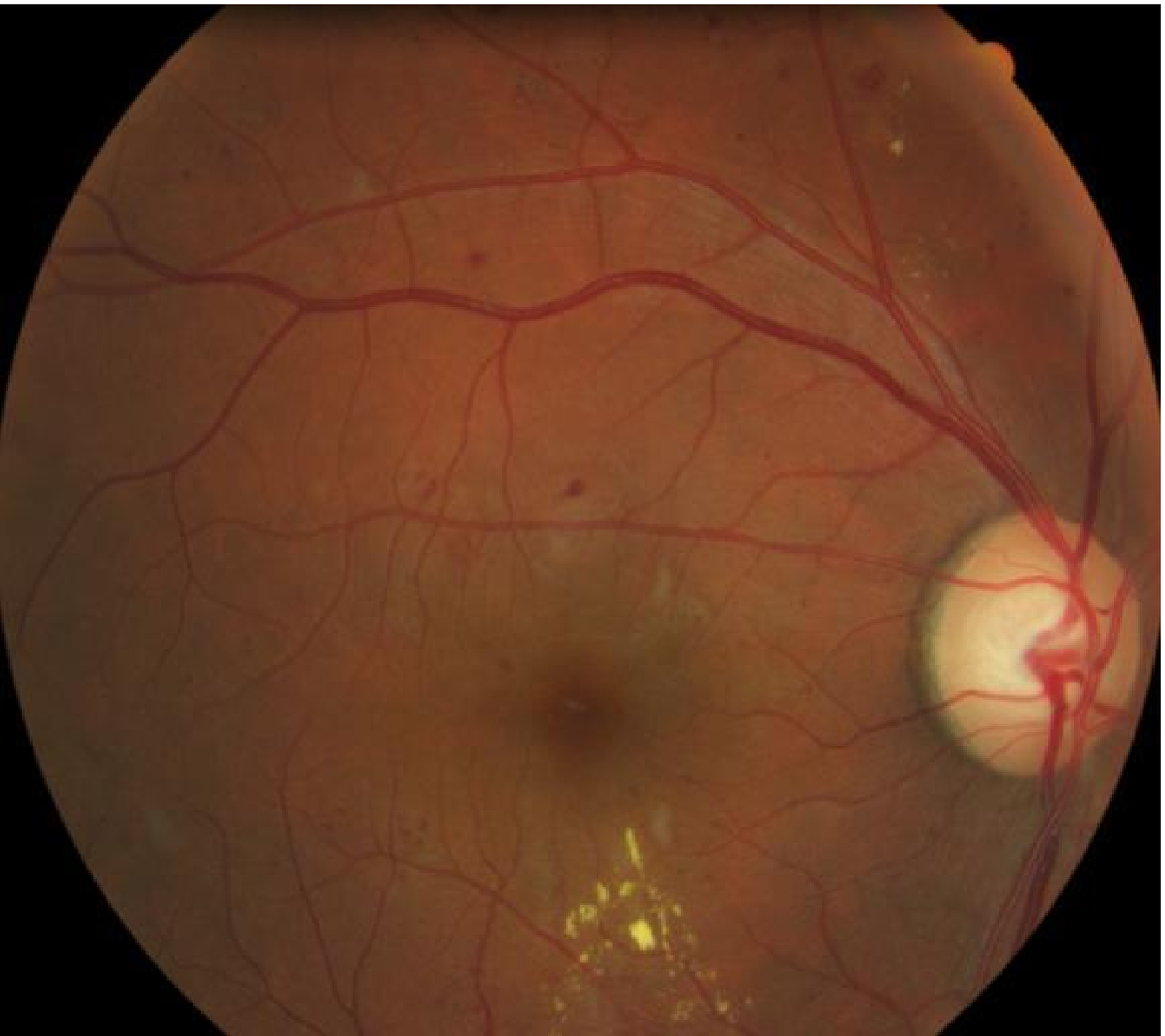}}
\subfigure[] {\label{cropping_a}\includegraphics[width=1in]{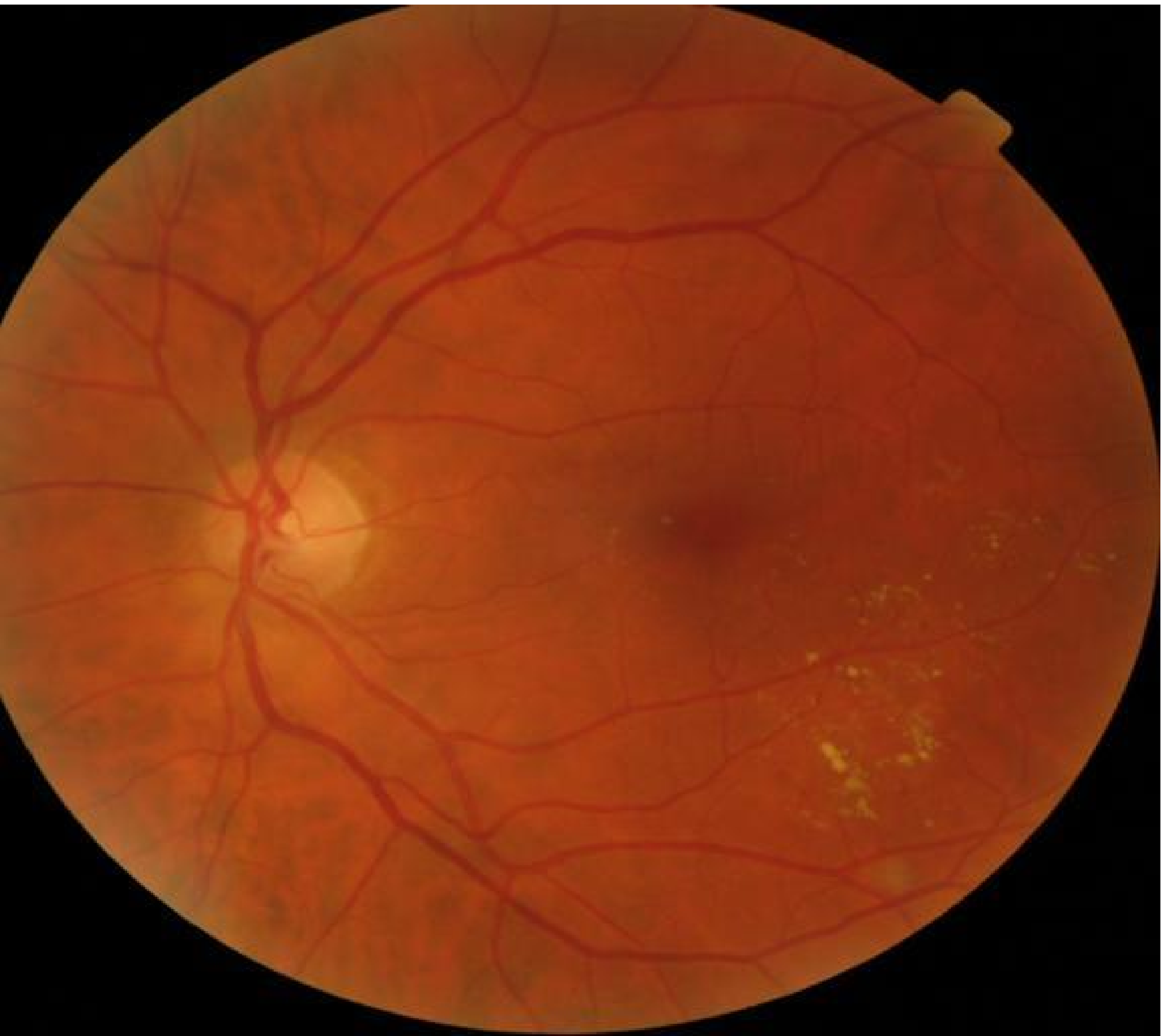}}
\subfigure[] {\label{cropping_a}\includegraphics[width=1in]{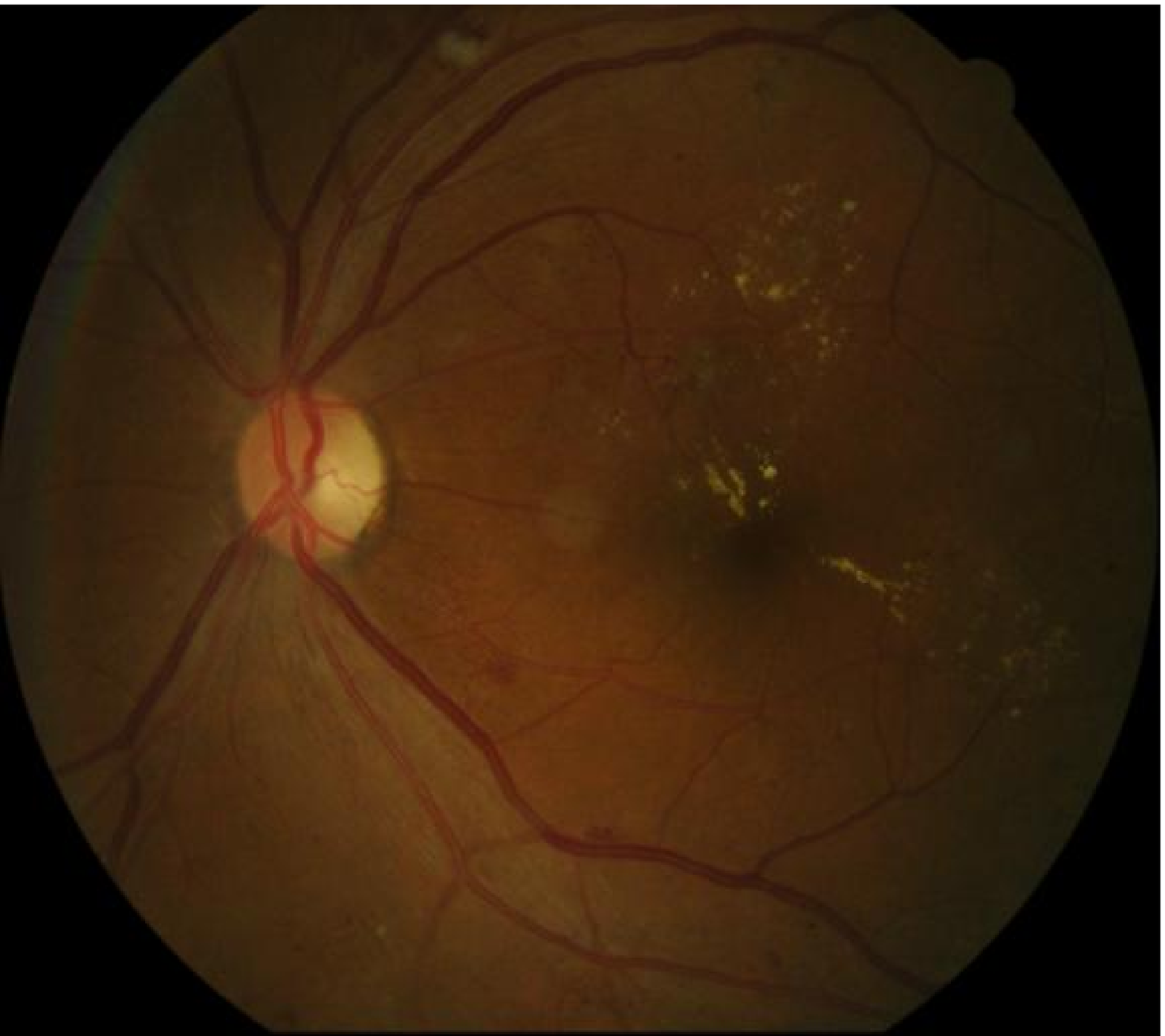}}

\subfigure[] {\label{cropping_a}\includegraphics[width=1in]{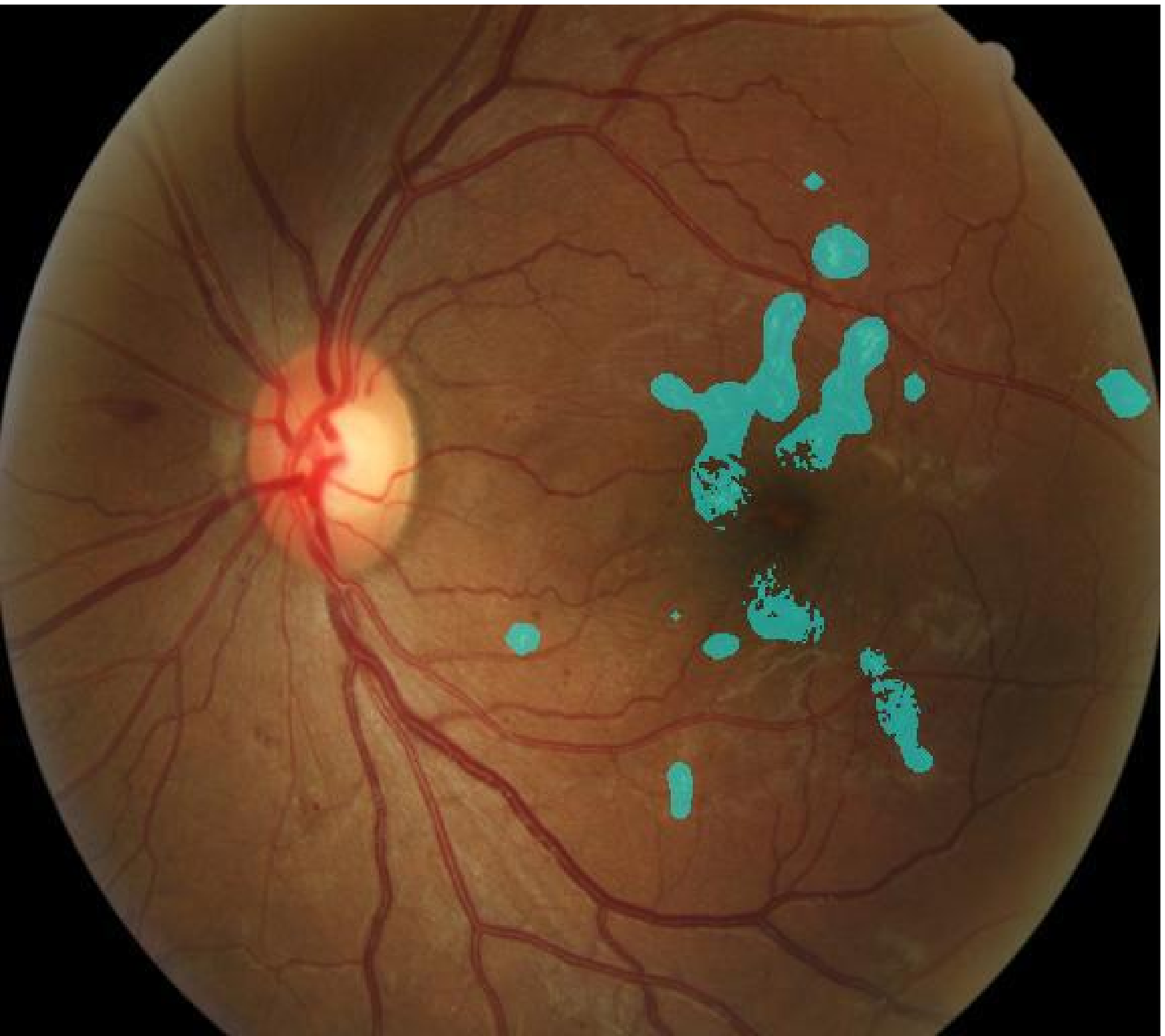}}
\subfigure[] {\label{cropping_a}\includegraphics[width=1in]{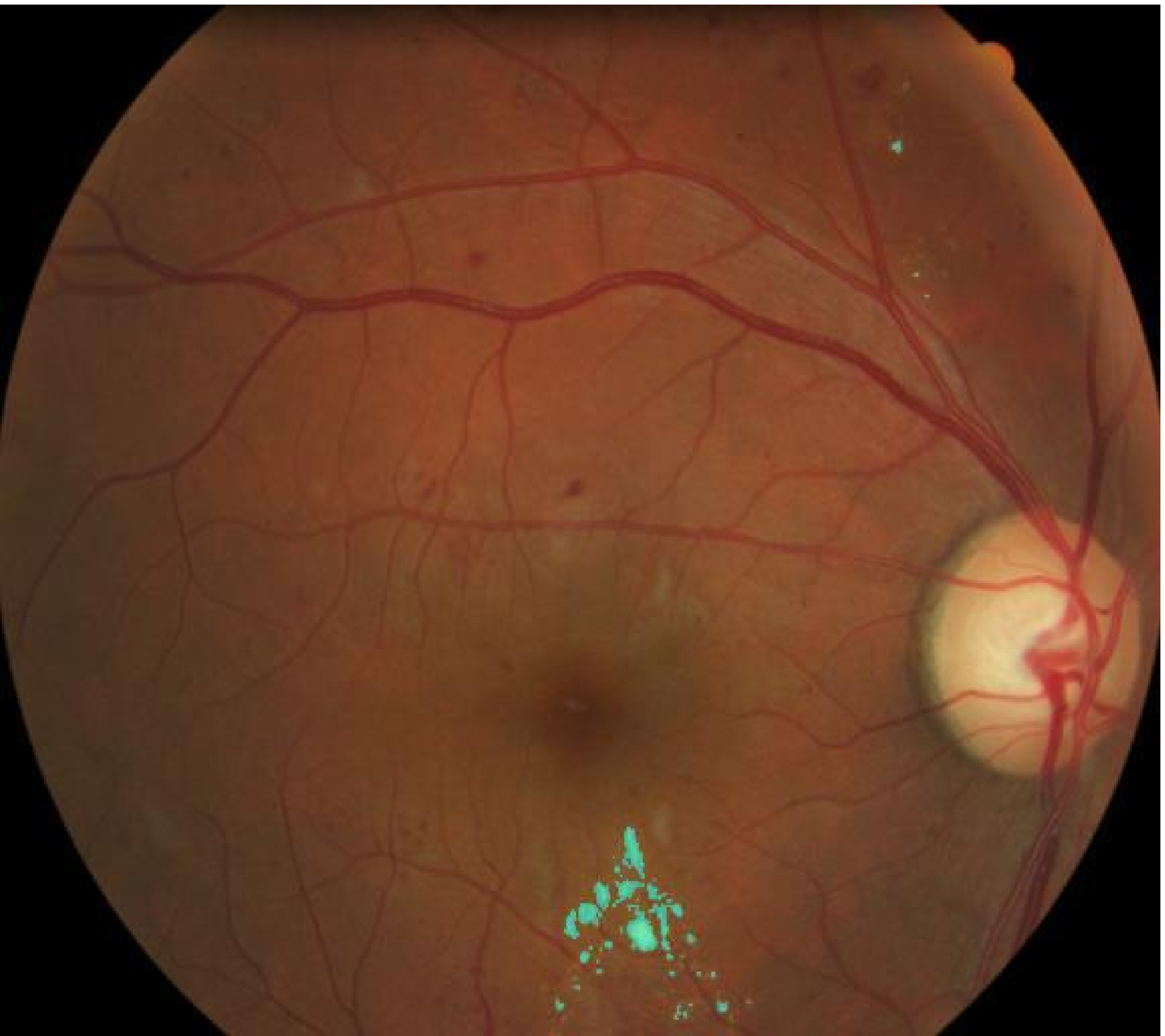}}
\subfigure[] {\label{cropping_a}\includegraphics[width=1in]{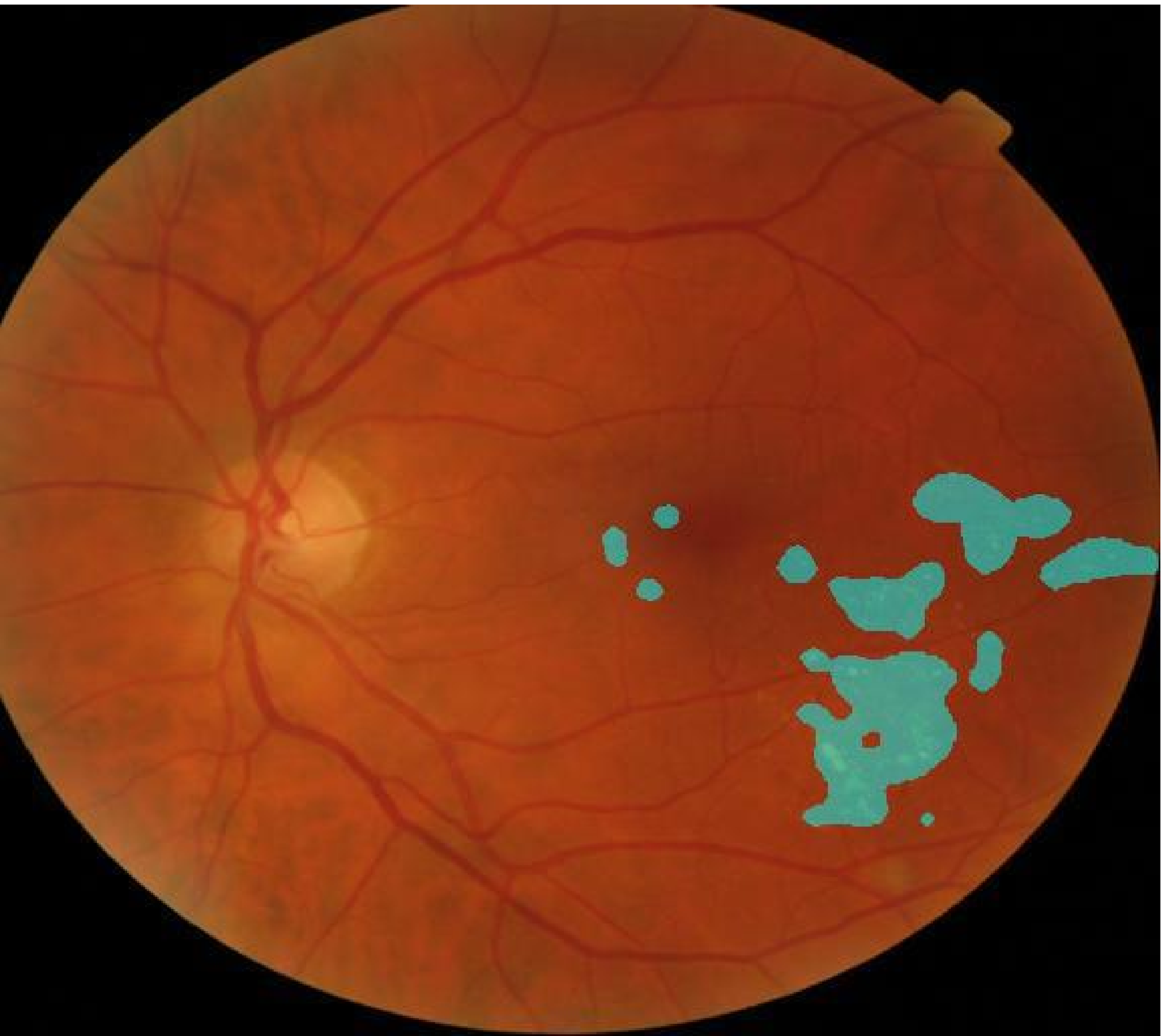}}
\subfigure[] {\label{cropping_a}\includegraphics[width=1in]{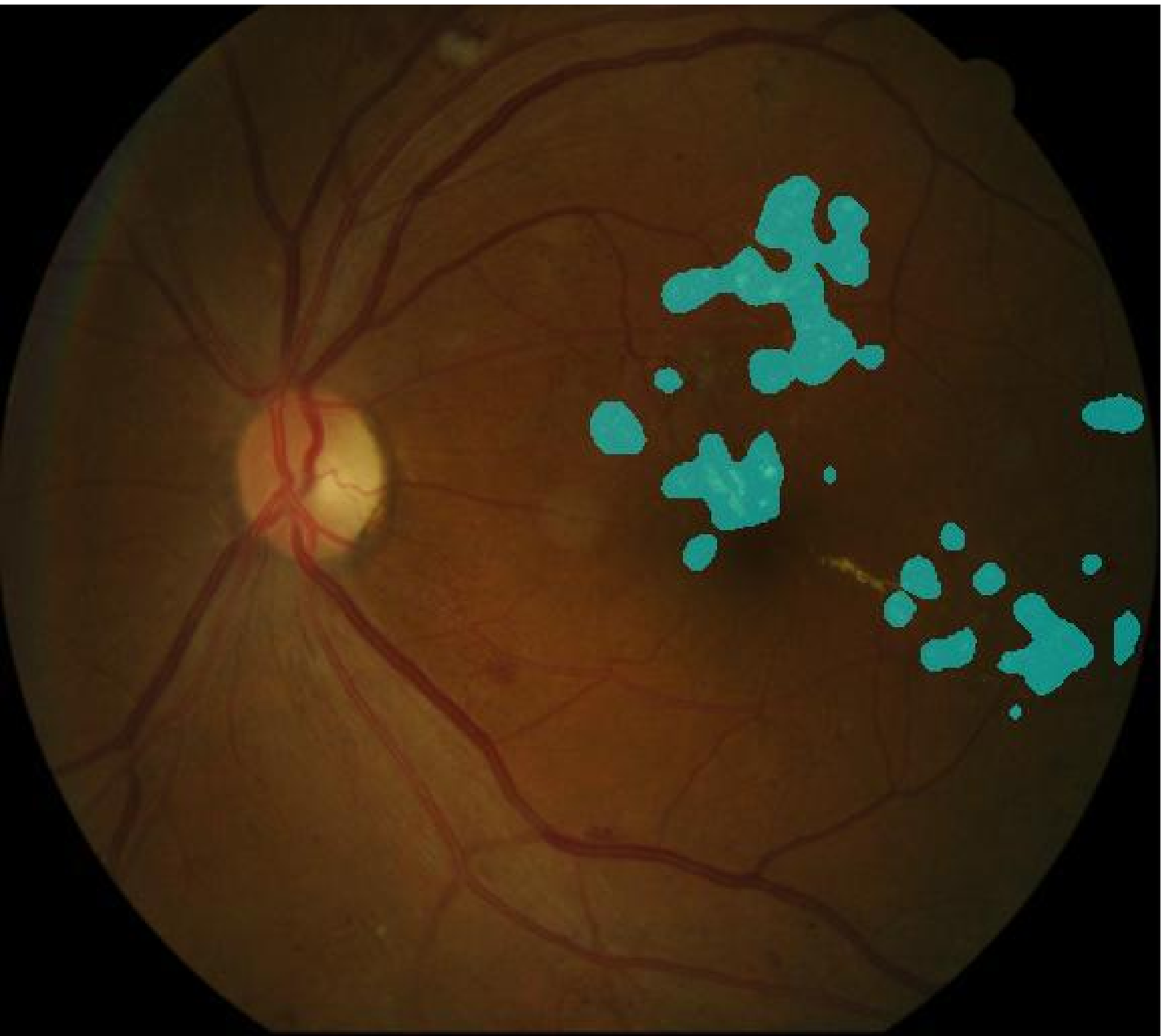}}

\subfigure[] {\label{cropping_b}\includegraphics[width=1in]{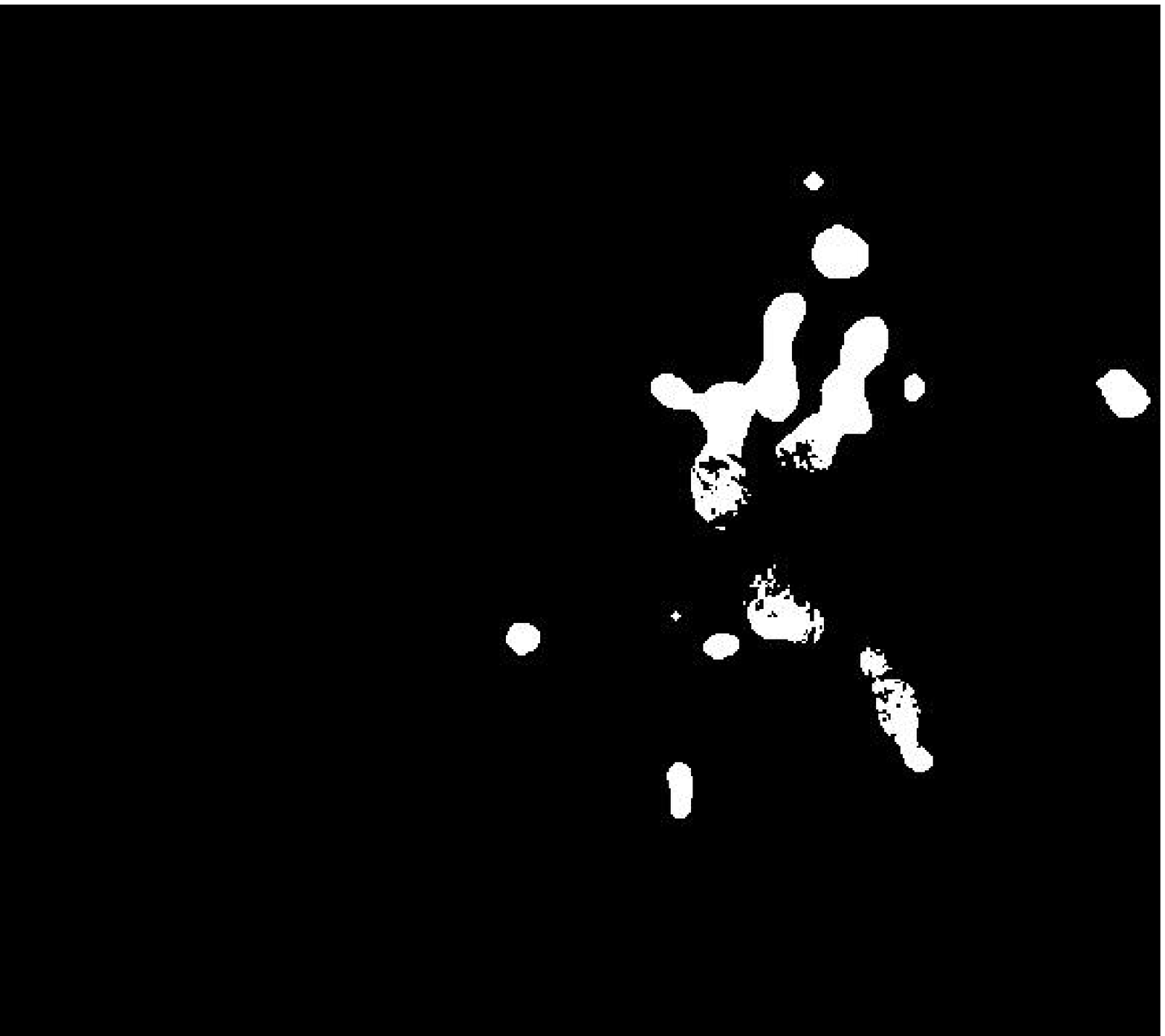}}
\subfigure[] {\label{cropping_b}\includegraphics[width=1in]{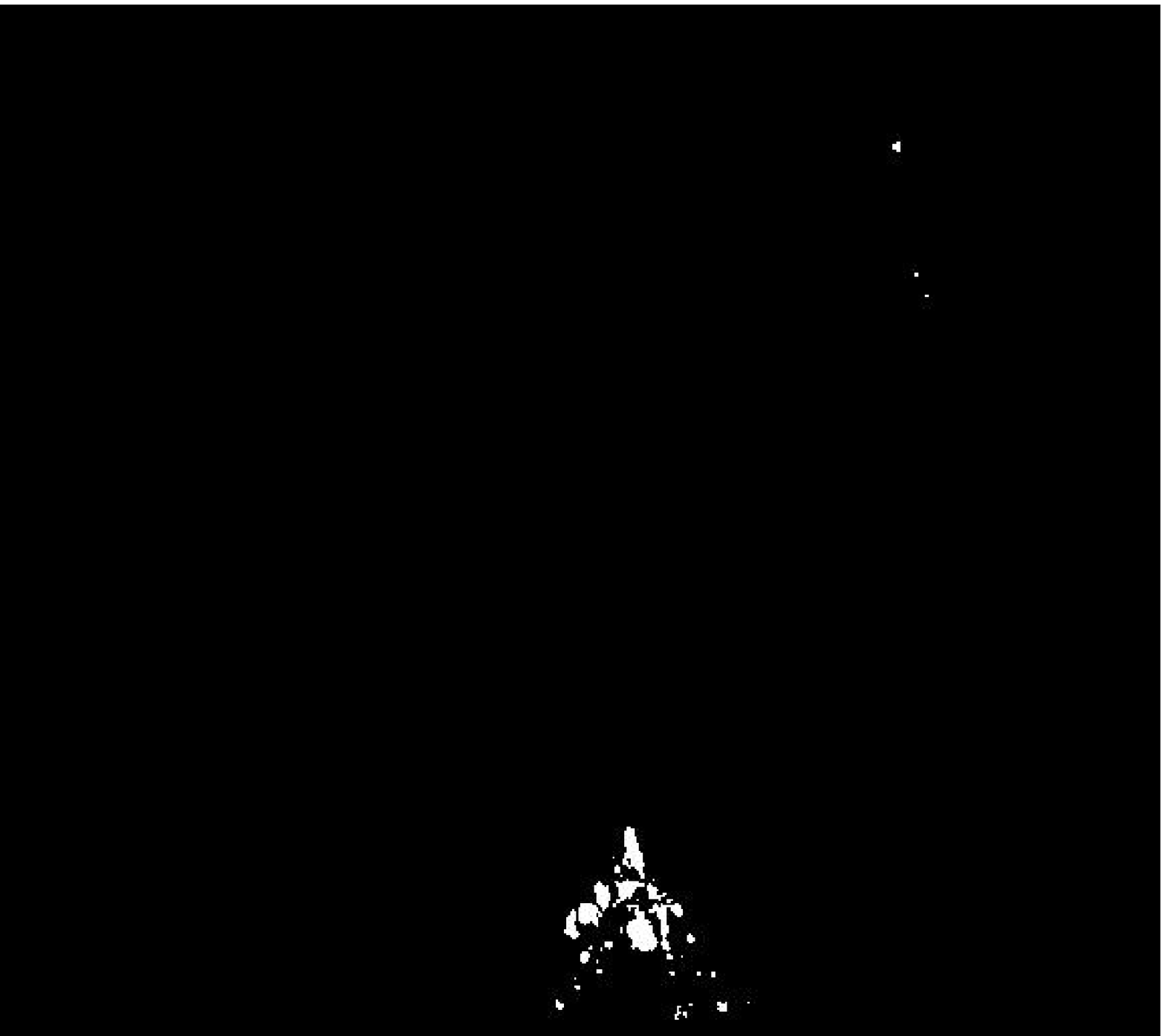}}
\subfigure[] {\label{cropping_b}\includegraphics[width=1in]{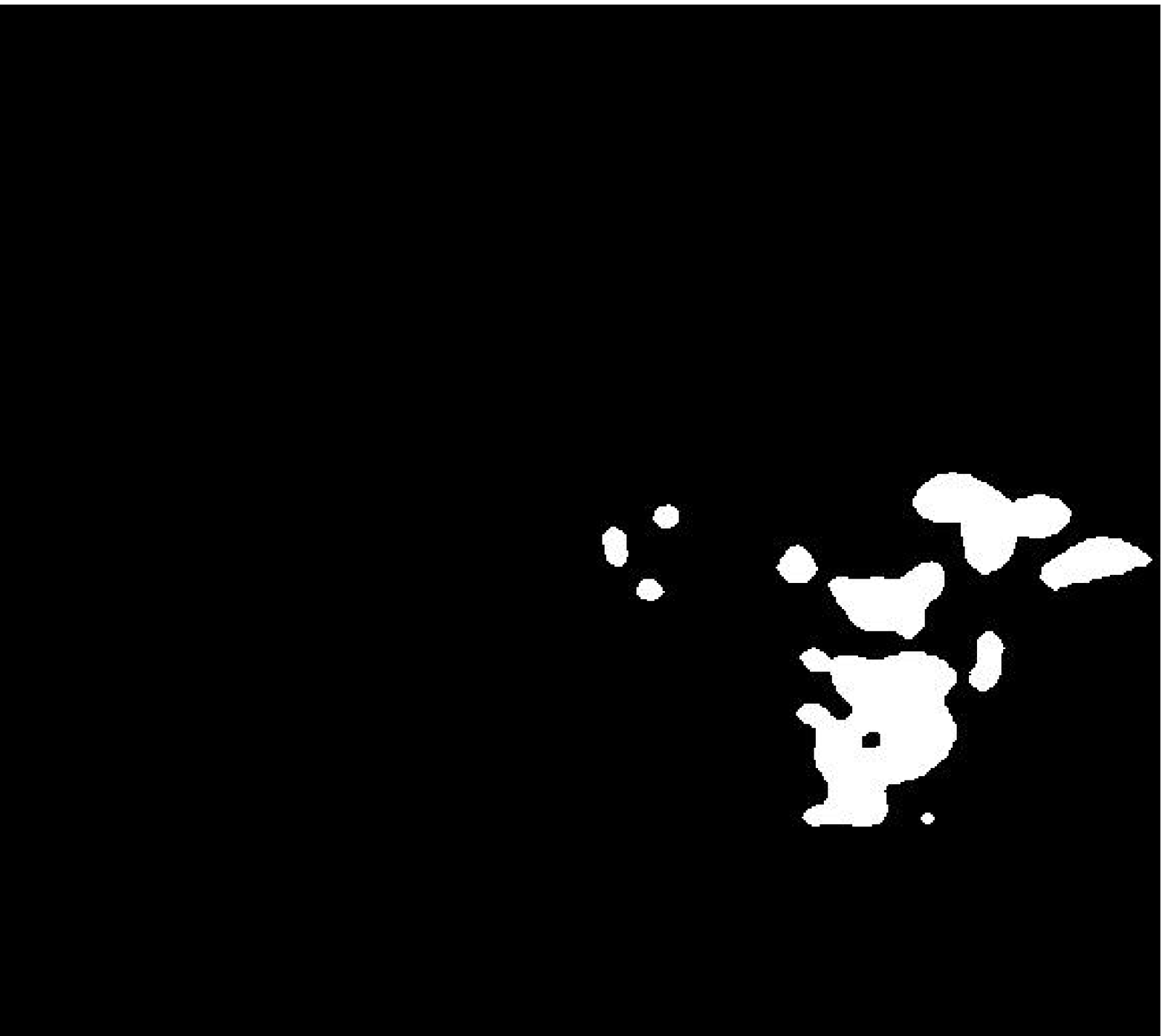}}
\subfigure[] {\label{cropping_b}\includegraphics[width=1in]{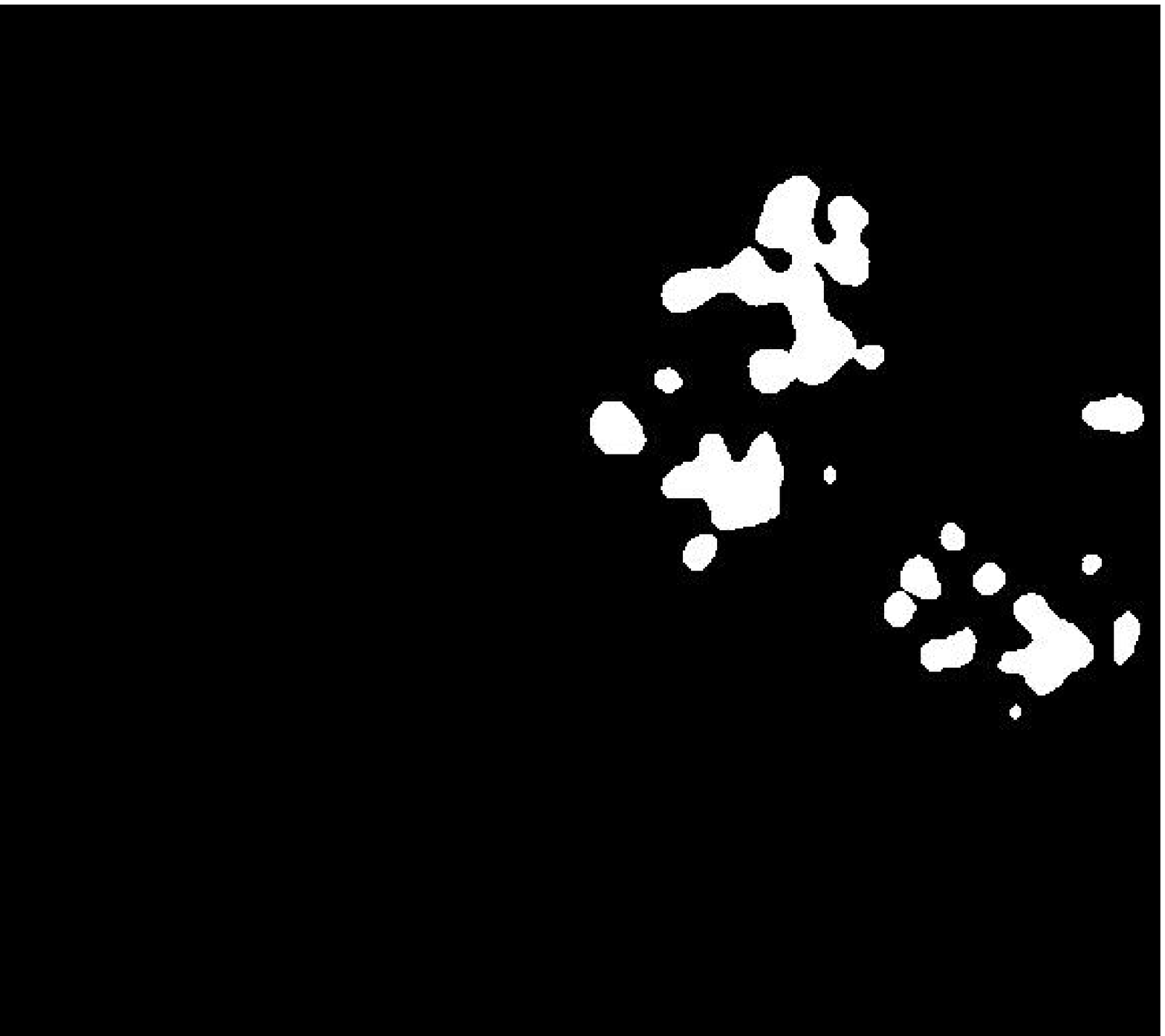}}
\caption {E-ophtha database results (a) Row 1 showing the original image (b) Row 2 showing the segmented overlay (c) Row 3 showing the result of segmented image}\label{visual}
\end{figure*}

\section{\bf \textsc{Conclusion}}
In this paper, we have proposed a supervised model of residual convolutional neural network for automatic segmentation of exudates symptom using RTC-Net architecture. From clinical perspective, extraction of exudates is the key step in the designing and development of computer-based diagnostic systems for ophthalmic conditions like Macular edma and diabetic retinopathy. The proposed method does not require any pre-processing or post-processing. The experimental results show that the proposed system achieves excellent results in terms of accuracy, sensitivity, and specificity for the same data set used previously. The RTC network trained on E-optha dataset with $60$ train images and $22$ test images provides an accuracy of 99\% with a sensitivity of 92\%. When  tested on the other two public datasets DiaReTDB1 and HEI-MED database, it provides an accuracy of 98\% and 98\% respectively with the sensitivity of 95\% and 97\% without any post-processing.

\section{\bf \textsc{Acknowledgment}}
In particular, the authors would like to thank E-Ophtha,  HEI-MED and DIARETDB databases management team for providing retinal images with manual annotations with tools. This database can been downloaded from the public links available.

\bibliographystyle{IEEEtran}      % basic style, author-year citations
\bibliography{References}   % name your BibTeX data base

\end{document}